\documentclass[twocolumn]{aastex631}

\newcommand{\mbh}{$M_{\rm BH}$}

\newcommand{\hst}{{\it HST}}
\newcommand{\msun}{$M_{\odot}$}

\submitjournal{ApJ}

\shorttitle{The BLR of NGC\,4151}
\shortauthors{Bentz et al.}

\begin{document}

\title{The Broad Line Region and Black Hole Mass of NGC\,4151}

\author[0000-0002-2816-5398]{Misty C.\ Bentz}
\affiliation{Department of Physics and Astronomy,
		 Georgia State University,
		 Atlanta, GA 30303, USA}
\email{bentz@astro.gsu.edu}

\author[0000-0002-4645-6578]{Peter R.\ Williams}
\affiliation{Department of Physics and Astronomy, 
    University of California, 
    Los Angeles, CA 90095, USA}

\author[0000-0002-8460-0390]{Tommaso Treu}
\altaffiliation{Packard Fellow}
\affiliation{Department of Physics and Astronomy, 
    University of California, 
    Los Angeles, CA 90095, USA}

\begin{abstract}

We present a reanalysis of reverberation-mapping data from 2005 for the Seyfert galaxy NGC\,4151, supplemented with additional data  from the literature to constrain the continuum variations over a significantly longer baseline than the original  monitoring program.  Modeling of the continuum light curve and the velocity-resolved variations across the H$\beta$ emission line constrains the geometry and kinematics of the broad line region (BLR).  The BLR is well described by a very thick disk with similar opening angle ($\theta_o\approx57\degr$) and inclination angle  ($\theta_i\approx58\degr$), suggesting that our sight line towards the innermost central engine skims just above the surface of the BLR.  The inclination is consistent with constraints from geometric modeling of the narrow line region, and the similarity between the inclination and opening angles is intriguing given previous studies of NGC\,4151 that suggest BLR gas has been observed temporarily eclipsing the X-ray source. The BLR  kinematics are dominated by eccentric bound orbits, with $\sim10$\% of the orbits preferring near-circular motions.  With the BLR geometry and kinematics constrained, the models provide an independent and direct black hole mass measurement of $\log M_{\rm BH}/M_{\odot} = 7.22^{+0.11}_{-0.10}$ or $M_{\rm BH}=1.66^{+0.48}_{-0.34}\times10^7$\,\msun, which is in good agreement with mass measurements from stellar dynamical modeling and gas dynamical modeling.  NGC\,4151 is one of the few nearby broad-lined Seyferts where the black hole mass may be measured via multiple independent techniques, and it provides an important test case for investigating potential systematics that could affect the black hole mass scales used in the local Universe and for high-redshift quasars.

\end{abstract}

%% https://astrothesaurus.org


\section{Introduction} %\label{sec:intro}

Reverberation mapping \citep{cackett21} is one of only a few methods that are able to directly constrain the mass of a supermassive black hole through its gravitational effects on luminous tracers (stars or gas).  Most methods that directly constrain black hole mass depend on spatial or angular resolution.  The mass of Sgr A* in the Galactic Center has been determined from decades of monitoring the proper motions of individual stars \citep{ghez00,genzel00,ghez08}.  Spatially resolved water maser clouds allowed accurate measurement of the mass of the central black hole in the nucleus of NGC\,4258 \citep{miyoshi95}.  And dynamical modeling of spatially resolved stellar and gas kinematics have produced a collection of over 100 black hole mass measurements for galaxies in the nearby Universe ($D\lesssim 100$\,Mpc; see the review by \citealt{kormendy13}).  In a few special cases, VLT-GRAVITY has been able to push beyond these typical distance limitations.  By combining 10-m class telescopes as an interferometer,  GRAVITY has successfully probed photoionized gas at sub-pc angular resolutions in the nuclei of active galaxies IRAS\,$09149-6206$ ($z=0.056$; \citealt{gravity20}) and 3C\,273 ($z=0.158$; \citealt{gravity18}) and determined constraints on their black hole masses.

In contrast, reverberation mapping relies on time resolution, using light echoes to probe the physical arrangement and conditions of photoionized gas around an accreting supermassive black hole.  With no angular resolution limit, reverberation mapping may be applied to active galaxies at any distance (e.g., \citealt{kaspi07,hoormann19,grier19,williams21a,williams21b}).  Results from reverberation mapping studies provide scaling relationships that allow quick estimation of large numbers of AGN black hole masses (e.g., \citealt{shen11}), permitting  studies of black hole growth and evolution as a function of lookback time.

In effect, there are two black hole mass scales currently in use and assumed to be equivalent: one based on the results of stellar and gas dynamical modeling in mostly early type galaxies within $D\approx100$\,Mpc, and one based on reverberation mapping results for active galaxies at larger distances.  All black hole mass measurement techniques include inherent uncertainties and and potential systematic biases (cf.\ \citealt{graham11,peterson10,kormendy13}), and at the moment it is not clear that dynamical modeling and reverberation mapping give the same results for the same black holes.  The Event Horizon Telescope results for P\={o}wehi, the nuclear black hole in M87, highlighted the need to compare direct black hole mass measurements: the mass derived from modeling of the interferometry data agreed with a previous measurement from stellar dynamics but disagreed with a gas dynamics measurement \citep{eht19}.  But with bright AGNs being rare within the volume that allows for high angular resolution in galaxy nuclei, there are few opportunities to directly compare masses from reverberation mapping and stellar or gas dynamics.

NGC\,4151 is one of the nearest broad-lined AGNs at $z=0.0033$, and thus one of only a handful of AGNs where the black hole mass may be measured using multiple independent techniques.  Its bright nuclear emission was first described by \citet{campbell18} based on observations collected, in part, by Dr.\ Heber Curtis, and computations likely conducted by unpaid assistant\footnote{\href{https://150w.berkeley.edu/celestial-observers-first-sixteen-berkeley-women-doctoral-graduates-astronomy-1913-1952}{https://150w.berkeley.edu/celestial-observers-first-sixteen-berkeley-women-doctoral-graduates-astronomy-1913-1952}} Miss Adelaide M.\ Hobe, who is credited with having carried out ``the major part of the computations'' in their study of bright line nebulae.  \citeauthor{campbell18} noted that NGC\,4151 had a spectrum resembling that of NGC\,1068 as described by \citet{fath09}.  Additional objects with similar properties were soon discovered, and \citet{seyfert43} conducted the first detailed investigation of so-called extragalactic nebulae with bright nuclear emission, including NGC\,4151.  We now know these objects as Seyfert galaxies. 

Variability on short timescales has come to be recognized as another typical characteristic of AGNs, including Seyfert galaxies. \citet{bahcall72} noted that flux variability from a central ionizing source would cause variations in surrounding photoionized gas in AGNs and some novae.  This idea was then developed into a framework for mapping out the geometry and kinematics of the broad line region (BLR) in AGNs by \citet{blandford82}, a technique they dubbed ``reverberation mapping''.

As a prototypical Seyfert galaxy with strong historic variability (cf.\ the 110 year light curve presented by \citealt{oknyanskij16}) and observed rapid variability in its bright nuclear spectral lines \citep{cherepashchuk73,antonucci83,bochkarev84} NGC\,4151 was a natural target for some of the first reverberation mapping studies \citep{peterson88,clavel90,ulrich91,maoz91}.  However, early spectroscopic monitoring programs were generally undersampled in the temporal domain because of an expectation from photoionization models that the BLR was an order of magnitude too large \citep{peterson85}.  This size problem was eventually solved by replacing single zone photoionization models with models that included gas covering a range of temperatures and densities, such as the LOC Model \citep{baldwin95}.  

Around the same time, it became clear that reverberation mapping could constrain the masses of the central black holes in these AGNs \citep{peterson99,peterson00a}.  By combining the average time delay for a broad emission line with its Doppler-broadened width, the black hole mass could be determined modulo a scaling factor that included important details such as the inclination angle at which we view the system.  As a stopgap, the use of a population-average scale factor $\langle f \rangle$ was introduced to bring black hole masses from reverberation mapping into broad agreement with the mass scale derived from stellar and gas dynamical modeling \citep{onken04}.  However, from the beginning, reverberation mapping was understood to be able to provide all the information needed to recover the full geometry and kinematics of the BLR, thus precluding the use of $\langle f \rangle$ and allowing for an independent and direct measurement of \mbh.

And so a series of intensive monitoring campaigns began in the early 2000s \citep{bentz06b,denney06} with the goals of improving the BLR measurements for objects that had previous measurements of poor quality or significantly undersampled data, and culminating in the acquisition of unambiguous velocity-resolved reverberation mapping data  (e.g., \citealt{bentz08,bentz09c,denney09c,grier12b}).  

 Constraining the details of the BLR from velocity-resolved reverberation data has been approached in two ways: either through forward modeling that explores the potential parameter space of BLR geometries and kinematics (e.g., \citealt{pancoast11}), or through the ill-posed inverse approach in which the time delay distribution as a function of velocity across the emission line (a velocity-delay map) is reconstructed directly from the data (e.g., \citealt{horne94,skielboe15,anderson21}). The two approaches are complementary.  Recovery of a velocity-delay map relies on a smaller set of core assumptions and is able to capture the full level of detail present in the data.  However, the interpretation of a velocity-delay map is not straightforward and relies on comparison with models.  Forward modeling, on the other hand, begins with a larger set of key assumptions to construct a fully self-consistent framework.  The need to rely on some simplifying assumptions ensures that the models may not fully explore the level of detail available in the data.  But the strength of forward modeling is that the results are relatively simple to interpret.  

The superb data sets that have finally begun to be  acquired by reverberation mapping programs have allowed the BLR structure and kinematics to be explored in detail for a modest number of AGNs using the forward modeling approach \citep{pancoast14b,grier17,williams18,williams20,bentz21c,villafana22a}.  While the exact details vary from AGN to AGN, these studies have generally found that the BLR, as probed by the H$\beta$ emission line, is arranged in a thick disk geometry that we are viewing at low to moderate inclination.  The kinematics are generally dominated by rotation that may also include a contribution from inflow or, in a few cases, some outflow.  In the few instances where velocity-delay maps have been cleanly recovered \citep{bentz10b,skielboe15,horne21}, the interpretations are in general agreement with the results from forward modeling.

Here, we reanalyze the spectroscopic monitoring data of NGC\,4151 that were obtained in early 2005 \citep{bentz06b} as part of the push to achieve velocity-resolved reverberation mapping data.  Poor weather significantly shortened the duration of the program, and so the original goals of the program were scaled back to simply determining an accurate H$\beta$ time delay from well-sampled, albeit short, light curves.  In this work, we supplement the original observations with additional measurements from the literature, extending the temporal coverage of the continuum variations by an extra $\sim 100$ days.  We model the continuum light curve and  H$\beta$ emission-line profiles with the phenomenological modeling code {\tt CARAMEL} \citep{pancoast11,pancoast14a},  and provide constraints on the BLR geometry and kinematics in NGC\,4151 along with an independent and direct measurement of the black hole mass.

\section{Observations and Data Preparation} 

 The data analyzed in this work come from several sources.  The observations that provide the H$\beta$ spectra and the bulk of the continuum flux measurements were originally described by \citet{bentz06b}, and we provide a summary here.  Long slit spectroscopy was collected on a $\sim$nightly basis between 2005 February 27 and 2005 April 10 at the MDM 1.3m McGraw-Hill Telescope.  Two to four 1200\,s spectra of NGC\,4151 were collected each night for a total of 96 individual spectra.  Typical CCD reductions were applied in IRAF and XVista.  

For this work, we re-extracted the spectra from the calibrated 2D frames and trimmed the spectra while applying a common dispersion solution, so that each spectrum covered $4400-5700$\,\AA\ with a dispersion of 1.25\,\AA\,pix$^{-1}$.  In previous work with these data, the nightly spectra were averaged together at this point and a typical flux uncertainty of 2\% was assumed.  Here, we continued our analysis with each individual spectrum and we carried the error array along with the spectra through our analysis, capturing the higher signal-to-noise (S/N) in the emission lines compared to the continuum and the additional noise from night sky lines.  A typical spectrum achieved S/N=100 per pixel in the continuum.

Variable seeing and slit placement over the two months of observations induced small variations in the wavelength solution, resolution, and flux calibration of the individual spectra.  These variations were minimized  with the \citet{vangroningen92} scaling algorithm using the [\ion{O}{3}]\,$\lambda$4959 emission line as an internal calibration source (the [\ion{O}{3}]\,$\lambda$5007 line was saturated in several of our spectra and could not be used).  

\begin{figure}
    \epsscale{1.17} 
    \plotone{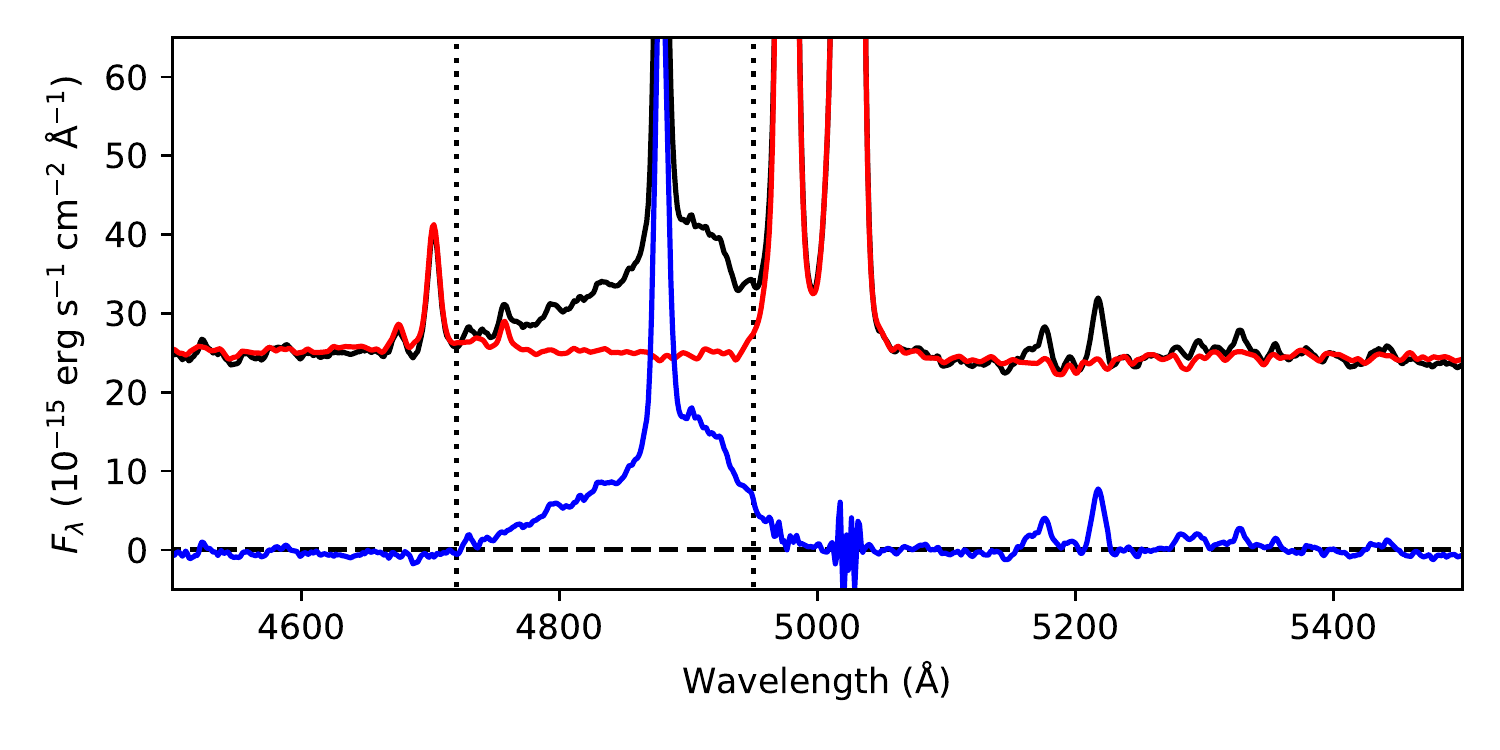}
    \caption{
    Example spectrum of NGC\,4151 (black) with the  model components that were subtracted from the spectrum (red) and the resulting isolated H$\beta$ emission (blue). The vertical dotted lines show the limits of the H$\beta$ emission that was used to constrain the dynamical models in this work.
    }
    \label{fig:spectrum}
\end{figure}

We then prepared the spectra for modeling by isolating the H$\beta$ emission through spectral decomposition.  Using the {\tt ULySS} package \citep{koleva09}, we modeled each spectrum with a host-galaxy component, a power law AGN continuum, and Gaussian components for the emission lines.  The host-galaxy component was selected from the \citet{vazdekis10} models derived from the MILES empirical stellar library, and was only allowed to vary in weight from spectrum to spectrum.  The H$\beta$ and [\ion{O}{3}] emission lines each required 3-5 Gaussians to capture their detailed shapes, and the many weak emission lines in the spectrum from \ion{Fe}{2} and other species were adequately modeled with single Gaussians.  As our goal here is to simply isolate the H$\beta$ emission, we do not attempt to interpret the model components beyond the goodness of fit that they provide to the observed spectra. Our process closely followed that of \citet{bentz21c}, where we first modeled the very high S/N mean spectrum, and then used the best-fit parameters for the mean spectrum as the initial parameters for the model of each individual spectrum.  Once a good fit was identified, we subtracted the host-galaxy and power law continuum components, as well as the [\ion{O}{3}], \ion{He}{2}, and other faint emission lines that were blended with H$\beta$.  Figure~\ref{fig:spectrum} displays an example spectrum in black, with the {\tt ULySS} model components that were ultimately subtracted shown in red and the isolated H$\beta$ spectrum in blue.

After isolating the H$\beta$ emission in each spectrum, we compared all spectra that were collected on a single night of observations.  In a few cases, weather conditions changed enough over the course of the observations ($\sim 1.0-1.5$\,hours) that the H$\beta$ profile in one spectrum deviated strongly from what was observed in the other spectra collected on that night.  We discarded three of the 96 spectra at this stage, and then averaged together the remaining spectra on each night to create 32 nightly spectra.  Finally, we cropped the nightly spectra outside of the range $4720-4950$\,\AA\ to focus solely on the H$\beta$ emission.  A small portion of the red wing of H$\beta$ was excluded by this region because of its position underneath the core of the bright [\ion{O}{3}]\,$\lambda$4959 emission line, where our spectral modeling process was sometimes unable to fully separate the two without leaving strong residuals behind.

Finally, we created the continuum light curve from our final sample of 93 re-extracted and scaled spectra by measuring the flux at rest-frame 5100\,\AA.  We supplemented this light curve with measurements from several additional sources to significantly extend the time baseline  as well as to improve the temporal sampling, when possible, beyond that provided solely by the MDM observations.  As described in the original analysis presented by \citet{bentz06b},  a few additional spectra were collected at the Crimean Astrophysical Observatory while the MDM observations were underway, and the continuum flux at 5100\,\AA\ from the Crimean spectra provided an additional eight measurements.  A further 24 measurements from $V-$band photometry that were collected as part of the MAGNUM project \citep{koshida14} were included, extending the continuum light curve approximately $100$\,days before the start of the spectroscopic monitoring (although with a coarse temporal cadence).  Finally, an additional three $V-$band measurements were added from photometry collected with the SARA telescopes \citep{roberts12}. 

\begin{figure}
    \epsscale{1.17} 
    \plotone{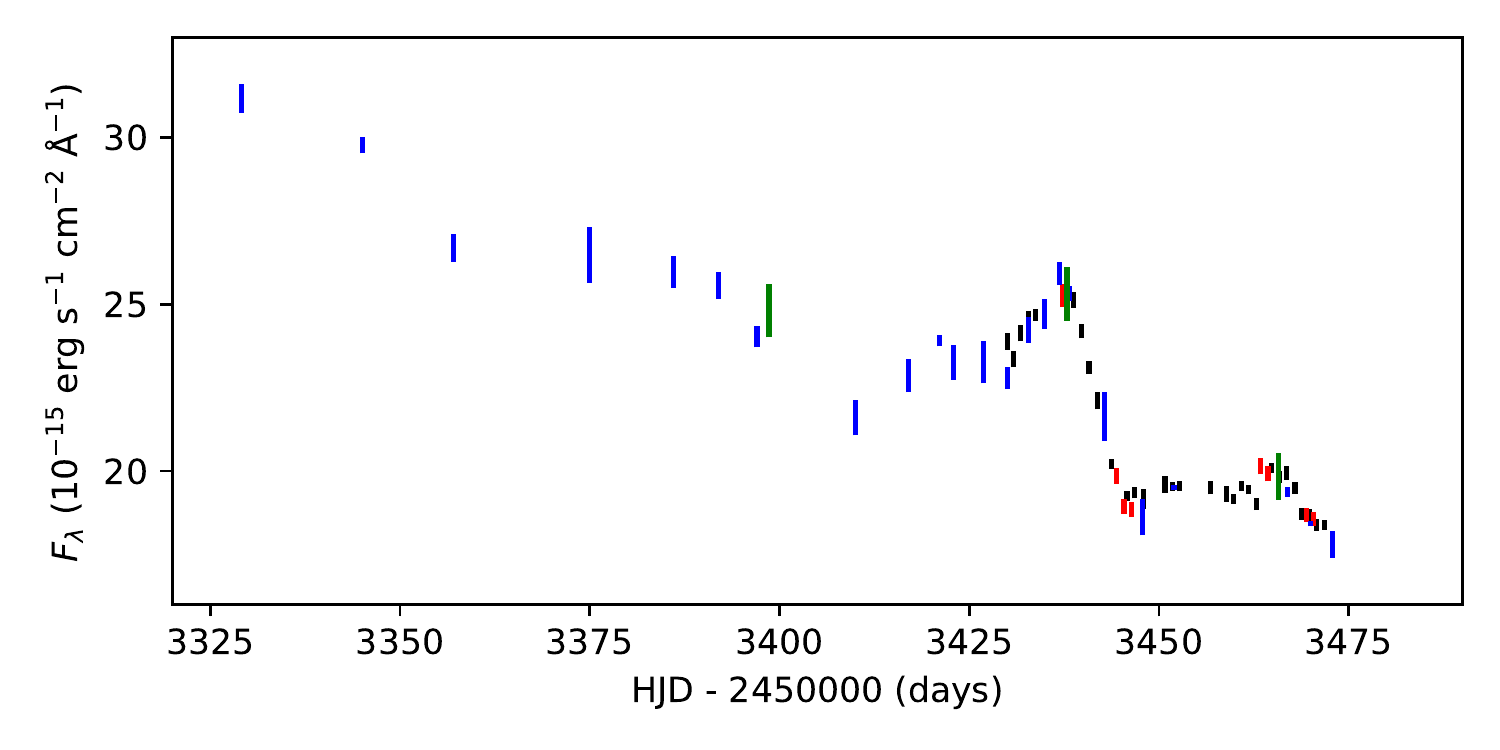}
    \caption{
    Continuum light curve for NGC\,4151 with data sets color coded as follows: spectroscopy from MDM (black), spectroscopy from Crimean Astrophysical Observatory (red), $V-$band photometry from the MAGNUM project (blue), and $V-$band photometry from the SARA telescopes (green).
    }
    \label{fig:cont_lc}
\end{figure}

All of these sets of measurements adopted different aperture sizes and are thus subject to different aperture losses and the inclusion of different amounts of host galaxy starlight.  Furthermore, there are bandpass differences between the spectroscopic measurements and the broad-band photometry.  To correct for these differences and calibrate each of the supplemental data sets to match the continuum fluxes measured from MDM spectroscopy, we follow the general procedure outlined by \citet{peterson91}.  We first identified  measurements that were made close in time ($\Delta t\lesssim 1$\,day) to MDM measurements.  For each supplemental data set, we fit a linear function to the close-in-time points from that observatory and from MDM, accounting for the uncertainties in each set of measurements.  The linear fit was then used to scale the supplemental dataset so that it matched the MDM measurements, and after all of the supplemental data sets were appropriately scaled to match the MDM measurements, they were merged together. Figure~\ref{fig:cont_lc} shows the four intercalibrated data sets with measurements from MDM spectroscopy in black, measurements from Crimean Astrophysical Observatory spectra in red, $V-$band photometry from the MAGNUM project in blue, and $V-$band photometry from the SARA telescopes in green.  The final continuum light curve was binned so that all measurements collected within $\Delta t=0.5$\,days were averaged together, providing 51 measurements over 143 days, with near-daily sampling during the final 42 days.  In comparison, the original continuum light curve analyzed by \citet{bentz06b} included 37 measurements covering just 41 days.

\section{Broad Line Region Models}

Modeling of the H$\beta$-emitting BLR was conducted with {\tt CARAMEL}, a phenomenological modeling code that is described in detail by \citet{pancoast14a}. {\tt CARAMEL} explores the BLR geometry and kinematics through the reverberation response across the velocity-resolved profile of a broad emission line as a function of time.  Here, we summarize the main components of the model.

The emissivity of the BLR in {\tt CARAMEL} is represented as a large collection of massless point particles that surround a massive black hole and are distributed in position and velocity space.  We note that these points should not be interpreted as physical objects, but rather as a Monte Carlo sampling of the line emissivity.  Continuum flux that is incident on a point particle is processed instantaneously, and the distribution of time delays from the BLR depends on the spatial distribution of the point particles, while the velocity distribution of the point particles gives the broad emission-line wavelength profile.

Radial and angular distributions are used to parameterize the spatial distribution of particles. The radial positions of the particles are drawn from a gamma distribution
\begin{equation}
    p(x|\alpha,\theta) \propto x^{\alpha - 1}\exp{\left( - \frac{x}{\theta} \right)}
\end{equation}
\noindent that is flexible enough to represent a Gaussian  ($\alpha>1$), an exponential ($\alpha=1$), or a cuspier profile ($0<\alpha<1$).  Experiments with different functional forms have shown that the results are insensitive to the choice of the gamma distribution.  The Schwarzschild radius, $R_s = 2GM/c^2$, plus an additional possible minimum radius $r_{\rm min}$ are used to shift the gamma distribution of particles away from the location of the black hole.  A change of variables is performed to assist with interpretation of the modeling results, which are given in terms of ($\mu$, $\beta$, $F$):
\begin{equation}
    \mu = r_{\rm min} + \alpha \theta,
\end{equation}
\begin{equation}
    \beta = \frac{1}{\sqrt{\alpha}},
\end{equation}
\begin{equation}
    F = \frac{r_{\rm min}}{r_{\rm min} + \alpha \theta},
\end{equation}
\noindent where $\mu$ is the mean radius, $\beta$ is the shape parameter, and $F$ is $r_{\rm min}$ in units of $\mu$.  The shifted gamma profile has a standard deviation given by $\sigma_r = \mu \beta (1-F)$. An outer radius of $r_{\rm out} = c \Delta t_{\rm data}/2$ truncates the BLR, where $\Delta t_{\rm data}$ is the time difference between the first point in the modeled continuum light curve and the first point in the emission-line light curve.  The  truncation at $r_{\rm out}$ assumes that the time baseline of the monitoring campaign is sufficiently long to track reverberation signals across the whole BLR.

The particle angular distribution is arranged in a disk with a thickness that corresponds to an opening angle $\theta_o$, where $\theta_o=0\degr$ is a thin disk and $\theta_o=90\degr$ is a sphere.  The disk inclination to the line of sight of an observer is set by $\theta_i$, where $\theta_i=0\degr$ is face on and $\theta_i=90\degr$ is edge on.  The distribution of particles as a function of depth within the disk sets the line emission strength.  For a single particle, the angle of displacement from the disk midplane is given by
\begin{equation}
    \theta_{d,N} = \arcsin (\sin \theta_o \times U^{1/\gamma})
\end{equation}
\noindent where $U$ is a random number that is drawn from a uniform distribution between 0 and 1.  The value of $\gamma$ ranges from 1 to 5, with a value of 1 corresponding to particles that are distributed uniformly throughout the thickness of the disk, while a value of 5 corresponds to clustered particles along the face of the disk, or emission that is preferentially from the outer skin of the BLR.  The  asymmetry parameter $\xi$ parameterizes  the amount of obscuration along the midplane of the disk, where $\xi \rightarrow 0$ causes the entire back half of the disk to be obscured and $\xi = 1$ has no midplane obscuration.  Finally  $\kappa$ provides a weight to each  particle 
\begin{equation}
    W(\phi) = \frac{1}{2} + \kappa \cos \phi
\end{equation}
where $W$ is the fraction of continuum flux that is radiated back towards the observer as line flux and $\phi$ gives the angle between the observer's line of sight to the source and the particle's line of sight to the source. The value of $\kappa$ ranges from $-0.5$ to $0.5$.  In the case of $\kappa=-0.5$, the particles preferentially emit back towards the ionizing source and an observer would see preferential emission from the far side of the disk.  Whereas for $\kappa=0.5$, the particles preferentially radiate away from the ionizing source and and observer would see preferential emission from the near side.

The particles are distributed in velocity space through  radial and tangential distributions.  Some fraction of the particles, $f_{\rm ellip}$, have near-circular orbits within the Keplerian potential of the central black hole with mass \mbh.  The remaining particles ($1-f_{\rm ellip}$) are either inflowing ($f_{\rm flow}<0.5$) or outflowing ($f_{\rm flow}>0.5$). However, these orbits may be highly eccentric and generally bound, or they may be unbound, as determined by the parameter $\theta_e$.  The possible values of the radial and tangential velocities define a plane, within which $\theta_e$ describes the angle of the velocity components towards the circular velocity and away from the escape velocity.  For $\theta_e =0\degr$, the  orbits are drawn from a Gaussian distribution centered on the escape velocity. For $\theta_e \rightarrow 90\degr$, the inflowing or outflowing orbits approach the parameter space occupied by near-circular orbits.   High values of $\theta_e$ indicate that inflowing or outflowing orbits are very nearly circular, while $\theta_e\approx45\degr$ indicates that most of the inflowing or outflowing orbits are highly eccentric but still bound, and low values of $\theta_e$ correspond to most particles being near the escape velocity and unbound. 

Included in the line-of-sight component of the velocity vector for each point particle is a contribution from macroturbulence, given by
\begin{equation}
    v_{\rm turb} = \mathcal{N} (0,\sigma_{\rm turb})|v_{\rm circ}|,
\end{equation}
\noindent where $v_{\rm circ}$ is the circular velocity and $\mathcal{N}(0,\sigma_{\rm turb})$ is a normal distribution centered on 0 and with standard deviation $\sigma_{\rm turb}$.

Parameterization of the spatial and velocity distributions of the particles allows an emission-line profile to be calculated for each continuum flux measurement, assuming that the continuum flux tracks the ionizing flux from a central point source. Included in the modeled emission-line profiles is a nonvariable narrow emission-line component, as well as a smoothing parameter to account for the small remaining differences in spectral resolution that arise from variable seeing conditions throughout the monitoring campaign. 

\begin{figure*}[h!]
    \epsscale{1.17} 
    \plotone{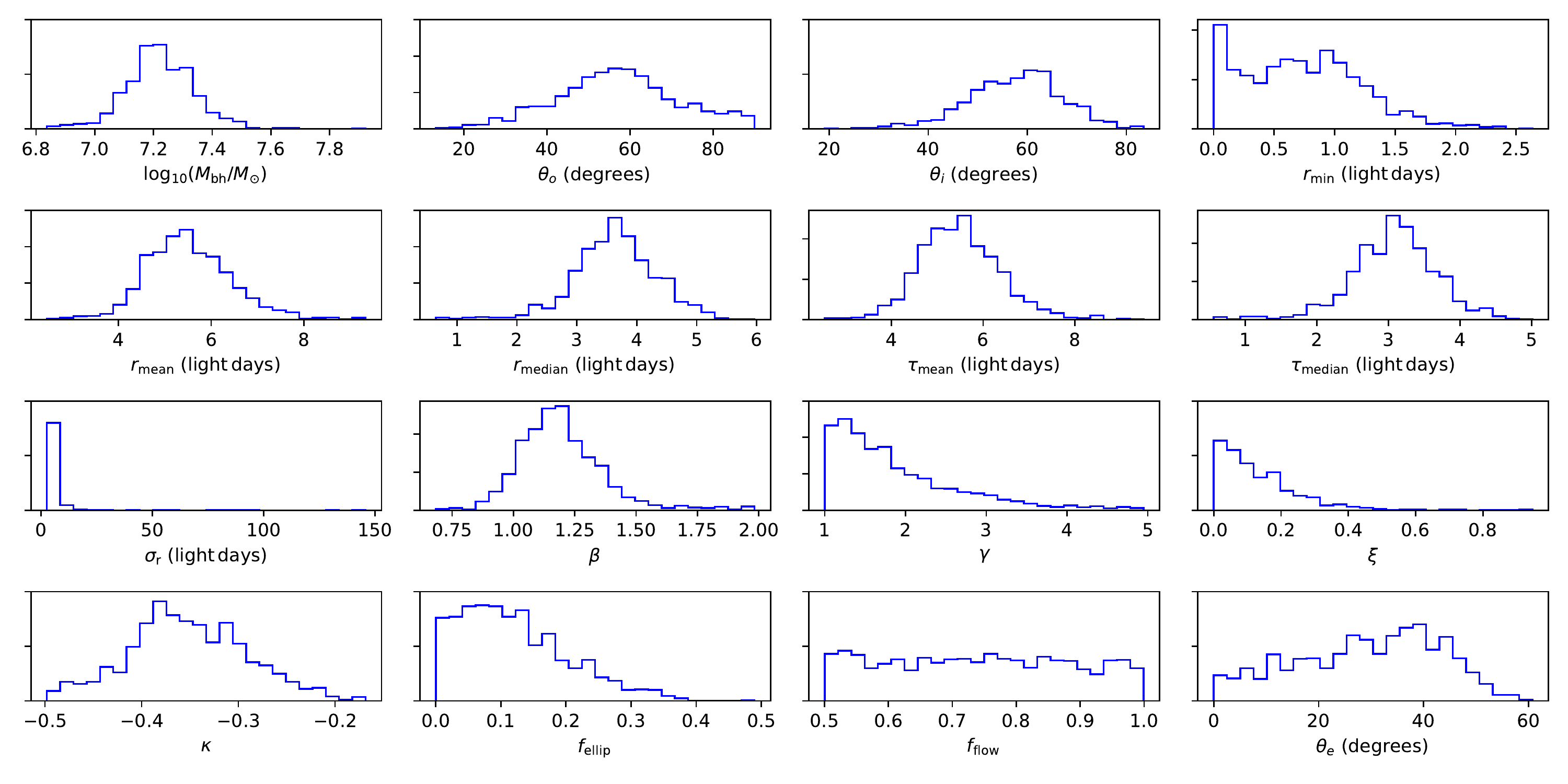}
    \caption{
    Posterior probability distributions for the H$\beta$ model parameters.
    }
    \label{fig:posteriors}
\end{figure*}

\begin{deluxetable*}{LlCC}[h!]
\renewcommand{\arraystretch}{1.1}
\tablecolumns{4}
%\tablewidth{0pt}
\tablecaption{Broad-line region model parameter values}
\tablehead{
\colhead{Parameter} & 
\colhead{Brief Description} & 
\colhead{H$\beta$} 
}
\startdata
\log_{10} (M/M_{\odot})         & Black hole mass                       & 7.22^{+0.11}_{-0.10}      \\
\theta_{o}\ \rm(degrees)        & Opening angle                         & 56.6^{+15.8}_{-14.3}        \\
\theta_{i}\ \rm(degrees)        & Inclination angle                     & 58.1^{+8.4}_{-9.6}        \\
r_{\rm min}\ \rm(light~days)     & Minimum radius of line emission       & 0.72^{+0.51}_{-0.55}     \\
r_{\rm mean}\ \rm(light~days)   & Mean radius of line emission          & 5.46^{+0.95}_{-0.79}     \\ 
r_{\rm median}\ \rm(light~days)  & Median radius of line emission        & 3.65^{+0.65}_{-0.63}    \\
\tau_{\rm mean}\ \rm(days)       & Mean time delay                       & 5.46^{+0.87}_{-0.78}     \\
\tau_{\rm median}\ \rm(days)     & Median time delay                     & 3.11^{+0.51}_{-0.55}     \\
\sigma_{r}\ \rm(light~days)      & Radial extent of line emission        & 5.74^{+1.64}_{-1.19}   \\
\beta                           & Shape parameter of radial distribution & 1.18^{+0.17}_{-0.14}     \\
\gamma                          & Disk face concentration parameter     & 1.67^{+0.98}_{-0.47}      \\
\xi                             & Transparency of the mid-plane         & 0.10^{+0.13}_{-0.07}      \\
\kappa                          & Cosine illumination function parameter & -0.36^{+0.06}_{-0.06}     \\
f_{\rm ellip}                   & Fraction of elliptical orbits         & 0.11^{+0.10}_{-0.07}      \\
f_{\rm flow}                    & Inflow vs.\ outflow                   & 0.74^{+0.17}_{-0.17}      \\
\theta_{e}\ \rm(degrees)         & Ellipse angle                         & 30.5^{+13.1}_{-17.7}     \\
\sigma_{\rm turb}               & Turbulence                            & 0.012^{+0.035}_{-0.009}   \\
r_{\rm out}\ \rm(light~days)     & Outer radius of line emission (fixed parameter) & 86             \\
T                               & Temperature or likelihood softening   & 2000                       
\label{tab:modelpars}
\enddata

\tablecomments{Tabulated values are the median and 68\% confidence intervals.}
\end{deluxetable*}

\vspace{-0.35in}
The continuum light curve must be interpolated in order to explore the range of possible time delays arising from the BLR and to properly compare the measured and the modeled emission line profiles.  {\tt CARAMEL} employs Gaussian processes to interpolate between continuum flux measurements as well as to extrapolate the continuum light curve beyond the start and end of the monitoring campaign, thus  extending the range of time delays that may be probed with the models.  The determination of the BLR model parameters includes the uncertainties on the Gaussian process model parameters, and so captures the effects of interpolating and extrapolating the continuum data within the quoted uncertainties.

For each model realization, 2000 individual point particles are used to represent the BLR.  The continuum light curve is interpolated and  emission-line profiles models are calculated for each epoch at which an emission-line measurement was acquired.  A Gaussian likelihood function compares the modeled spectra with the measured spectra and adjusts the model parameters accordingly. {\tt CARAMEL} utilizes a diffusive nested sampling code based on {\tt DNEST4} \citep{brewer18} to efficiently explore the parameter space of the models.  {\tt DNEST4} allows for the use of a likelihood softening parameter, or statistical temperature $T$, which has the effect of increasing the measurement uncertainties.  The likelihood softening parameter is able to account for underestimated measurement uncertainties and for the inability of the model to capture all of the complex and real details in the measurements.  After completing 10,000 model runs, the value of $T$ is determined in the post analysis by examining the distributions of the model parameters and choosing the largest value of $T$ for which the distributions remain smooth and generally unimodal.  To verify that convergence had been reached, we compared the values of the model parameters from the first half of the model runs to the values determined for the second half of the model runs, finding no significant difference between the parameters constrained from either half.

\subsection{Model limitations}

Exploring the parameter space of a BLR model requires quick and repeated calculations to compute many emission line time-series. Due to this constraint, the model must be simplified, and with these simplifications come limitations.

First, the model excludes some physics such as radiation pressure and photoionization processes. Including these would require additional assumptions about the ionizing continuum, which is generally not observable, and the BLR gas properties. Neglecting radiation pressure is standard in reverberation mapping studies as gravity is assumed to be the dominant force affecting BLR kinematics, especially in low Eddington ratio sources like NGC\,4151 \citep[$L_{\rm bol}/L_{\rm Edd} \sim 0.01-0.1$,][]{merritt22}. Since we do not include photoionization processes, however, it is critical to understand that the models investigate the BLR \textit{emissivity} distribution, and extreme care should be taken when using the model results to describe the BLR gas. For instance, a model with $\gamma=5$ indicates that the observed emissivity is concentrated at the ``skin'' of the thick BLR disk. This could mean there is no gas in the inner parts of the disk or that the emission from the inner parts is obscured, but our models cannot distinguish between the two scenarios.

Second, the model is parametrized in such a way that the maximum range of geometries can be described with the fewest number of free parameters. This requires the use of smooth, continuous functions to describe the BLR emissivity. Thus,  short-timescale fluctuations in the emission line profile, corresponding to short size-scale fluctuations in the BLR, cannot be modeled. 

Third, because photoionization processes are not included, the {\tt CARAMEL} model does not fit the absolute flux scale of the emission line but rather re-scales the continuum fluctuations so that the two scales match. Modeling the absolute fluxes would again require knowledge of the ionizing continuum as well as the physical properties of the BLR gas. By re-scaling the fluxes in this way, the emission line profile shapes can be fit and provide constraints on the BLR without making additional assumptions.

Despite these limitations, proper interpretation of the model---as a Monte Carlo approximation of the BLR kinematics and emissivity field---provides significant information about properties of the BLR that are of the most interest, such as the size, orientation, overall structure and kinematics, as well as the black hole mass. Repeat modeling of the same AGN over multiple observing campaigns demonstrates that parameters expected to remain constant are robust \citep[e.g., Arp\,151,][]{pancoast18}. Modeling of multiple emission lines from the same AGN find ionization stratification consistent with theory (NGC\,5548, \citealt{williams20}; NGC\,3783, \citealt{bentz21c}). Constraints on the BLR kinematics are consistent with those found from velocity-resolved RM and the maximum entropy method \citep[e.g.,][]{villafana22a}, and inclination angles are consistent in the few cases in which independent measurements are available (\citealt{grier17}; this work). Tests of simulated data also show reassuring results, in that the key properties of the BLR and the black hole mass are robustly recovered even when the input model is significantly different than what is assumed in {\tt CARAMEL} \citep{mangham19}.

\section{Results}

The median and 68\% confidence intervals for all of the H$\beta$ BLR model parameters in NGC\,4151 are listed in Table~\ref{tab:modelpars}, and Figure~\ref{fig:posteriors} shows the posterior probability function for each parameter.

\begin{figure}
    \epsscale{1.17} 
    \plotone{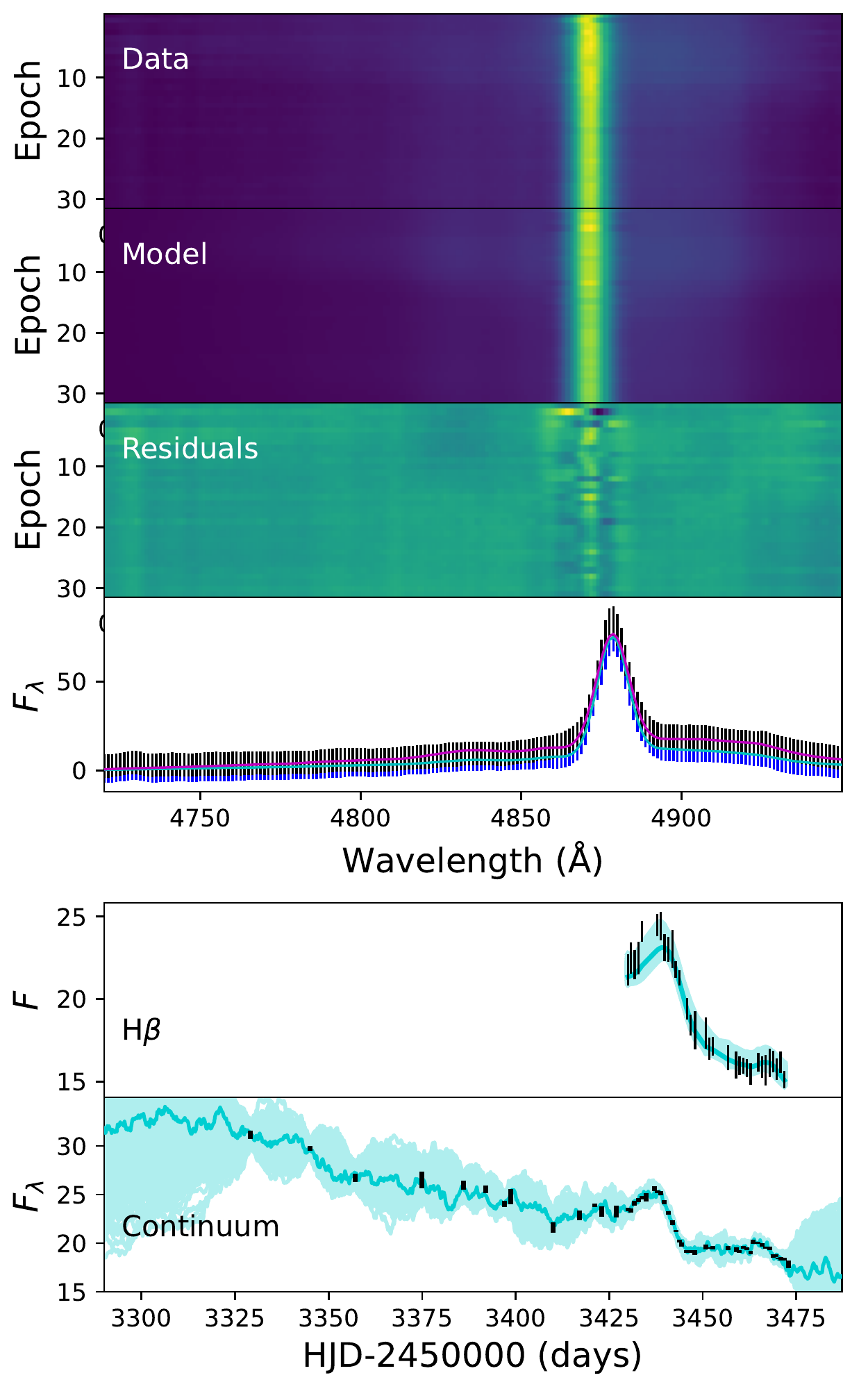}
    \caption{
    The three panels at the top display the data, one possible model, and residuals (data$-$model) for the H$\beta$ spectra.  Immediately below are a low flux spectrum (epoch 20; blue error bars) and a high flux spectrum (epoch 6; black error bars)  with model fits overlaid as the cyan and magenta curves, respectively. In the bottom two panels, the continuum and integrated H$\beta$ light curves are displayed as data points with model fits overlaid. The full ranges of the models are displayed in light turquoise with the example model corresponding to the top four panels overlaid in dark turquoise.  All uncertainties associated with H$\beta$ have been expanded by a factor of $\sqrt{T}=44.7$.  Flux densities ($F_{\lambda}$) are given in units of $10^{-15}$\,erg\,s$^{-1}$\,cm$^{-2}$\,\AA$^{-1}$ while integrated flux ($F$) is in units of $10^{-13}$\,erg\,s$^{-1}$\,cm$^{-2}$. Across the six panels, it is evident that most of the gross characteristics of the data are captured by the models.  
    }
    \label{fig:datamodel}
\end{figure}

The models required a likelihood softening of $T=2000$, or an increase in the uncertainties of a factor of $\sqrt{T}=44.7$. We note that the very high S/N in each epoch of spectroscopy considered here suggests that the statistical noise is negligible compared to modeling and systematic errors, thus requiring $T>>1$. Figure~\ref{fig:datamodel} displays the continuum and integrated H$\beta$ light curves as well as the observed emission-line profiles, with the model fits to all displayed as well. The H$\beta$ uncertainties have been expanded by $\sqrt{T}=44.7$, and the models capture the gross characteristics of the H$\beta$ profiles and variations quite well.

\begin{figure}
    \epsscale{1.17} 
    \plotone{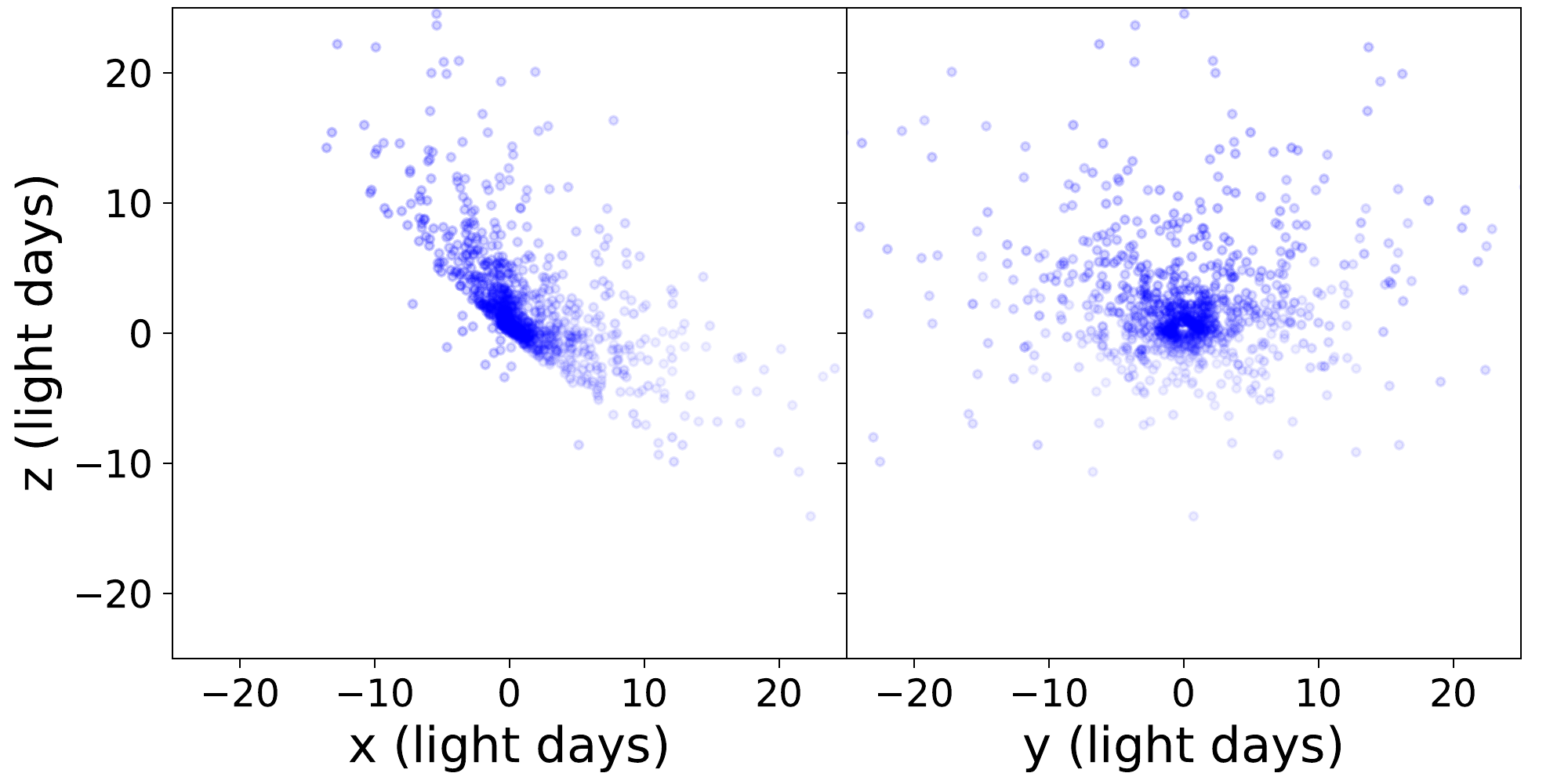}
    \caption{
    One representative geometric model for the H$\beta$ response in the broad line region of NGC\,4151, drawn from the posterior probability distribution.  The left panel is an edge-on view of the structure, with an Earth-based observer located on the +x axis, while the right panel shows the observer's view.  The transparency of each point in the image represents the relative response of the gas to continuum fluctuations, with more opaque points responsible for a stronger response. The far side of the disk displays a much stronger response than the near side, and the midplane of the disk is almost completely opaque. 
    }
    \label{fig:geometry}
\end{figure}

The geometry of the H$\beta$ emitting region in NGC\,4151 at the time of the observations is best represented by a very thick disk (opening angle $\theta_o=56.6^{+15.8}_{-14.3}$\,deg) inclined by $\theta_i=58.1^{+8.4}_{-9.6}$\,deg to our line of sight.  The disk has an inner minimum radius of $r_{\rm min}=0.72^{+0.51}_{-0.55}$\,lt-day, with a mean and median radius of $r_{\rm mean}=5.46^{+0.95}_{-0.79}$\,lt-day and $r_{\rm median}=3.65^{+0.65}_{-0.63}$\,lt-day, respectively.  The radial width of the emission is $\sigma_r=5.74^{+1.64}_{-1.19}$\,lt-day, and the radial distribution of the emission has a profile that is slightly more cuspy than an exponential ($\beta=1.18^{+0.17}_{-0.14}$).  The emission is distributed fairly uniformly through the disk, with a slight preference for stronger emission near the face of the disk ($\gamma=1.67^{+0.98}_{-0.47}$) and almost complete obscuration along the midplane ($\xi=0.10^{+0.13}_{-0.07}$).  Most of the line emission is preferentially directed back towards the central illuminating source ($\kappa=-0.36^{+0.06}_{-0.06}$), with the observer seeing a strong response from the far side of the disk and weak response from the front.  Figure~\ref{fig:geometry} displays a representative geometric model, drawn from the posterior probability distribution for the H$\beta$ emission-line response in NGC\,4151.

\begin{figure}
    \epsscale{1.17} 
    \plotone{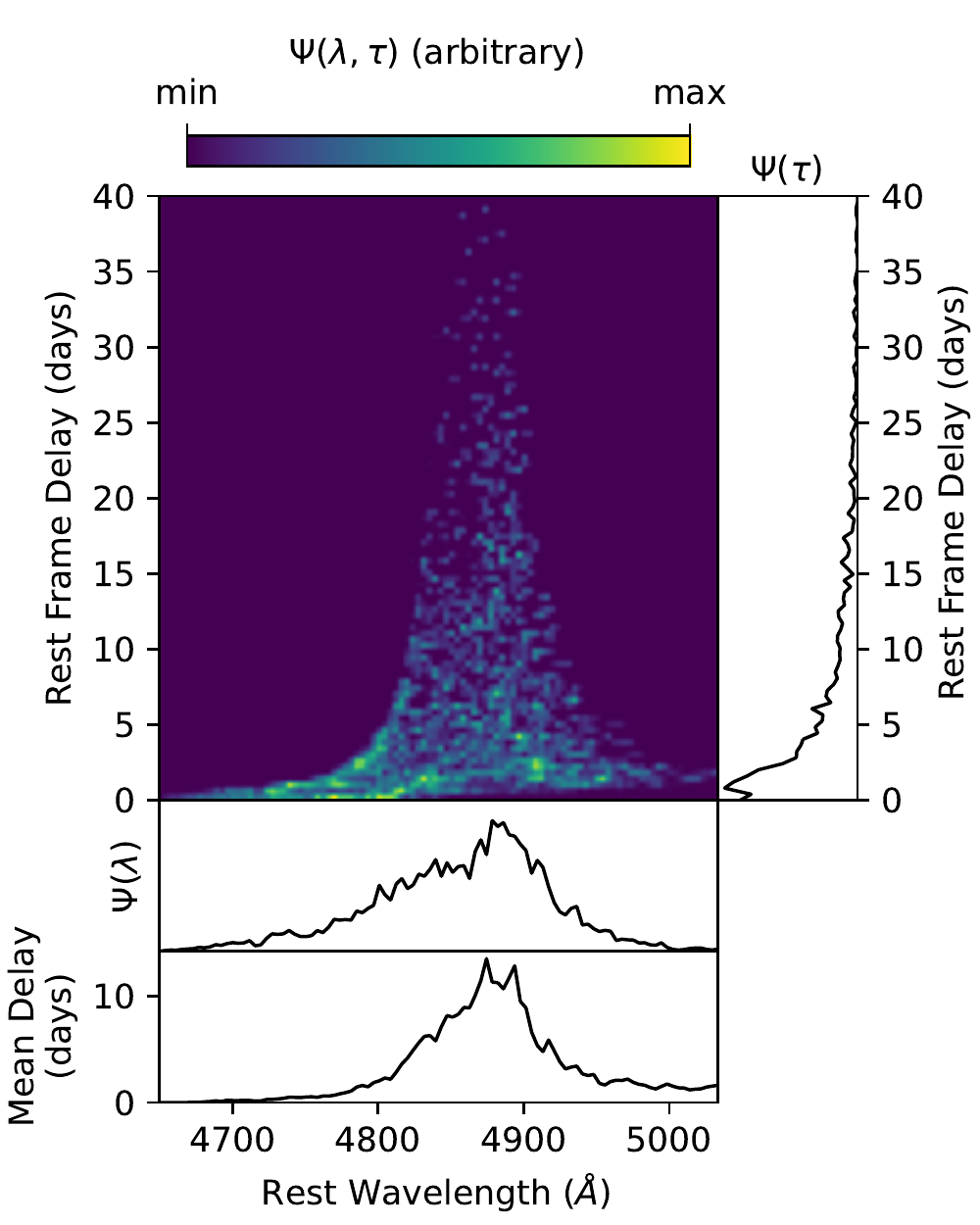}
    \caption{
    Transfer function $\Psi(\lambda,\tau)$ for the example H$\beta$ model displayed in Figure~\ref{fig:geometry}.  Integrating the transfer function over wavelength gives the one-dimensional lag profile $\Psi(\tau)$, which is shown on the right.  Integrating the transfer function over time delay gives $\Psi(\lambda)$, or the variable emission-line profile, which is shown immediately under the transfer function.  The bottom panel displays the average lag as a function of wavelength across the emission line. 
    }
    \label{fig:hbtransfer}
\end{figure}

The mean and median time delays associated with this geometry are $\tau_{\rm mean}=5.46^{+0.87}_{-0.78}$\,days and $\tau_{\rm median}=3.11^{+0.51}_{-0.55}$\,days.  The mean time delay agrees well with the H$\beta$ time delay reported by \citet{bentz06b} of $\tau_{\rm cent}=6.59^{+1.12}_{-0.76}$\,days.  Figure~\ref{fig:hbtransfer} shows a transfer function $\Psi(\lambda,\tau)$ for a representative model.  The transfer function depicts the strength of H$\beta$ responsivity as a function of velocity and time delay across the observed H$\beta$ line profile.

The black hole mass is found to be $\log_{10}(M/M_{\odot})=7.22^{+0.11}_{-0.10}$.  About 10\% of the orbits are near circular ($f_{\rm ellip}=0.11^{+0.10}_{-0.07}$), while the majority ($\sim90$\%) prefer outflow-like kinematics ($f_{\rm flow}>0.5$).  With $\theta_e=30.5^{+13.1}_{-17.7}$\,deg, most of these orbits are highly eccentric but still bound.  Finally, there is a very small kinematic contribution from turbulence ($\sigma_{\rm turb}=0.012^{+0.035}_{-0.009}$).

\section{Discussion}

Previous studies that model velocity-resolved reverberation mapping data have consistently found a preference for BLR geometries that resemble a moderately-inclined thick disk  \citep{brewer11,pancoast14b,grier17,williams18,bentz21c,villafana22a}.  The opening angle and inclination angle that we find here suggest that NGC\,4151 has the thickest and most highly inclined BLR disk among Seyfert 1s that have been studied with these methods \citep{villafana22b}. 

The inclination angle to our line of sight, $\theta_i=58.1^{+8.4}_{-9.6}$\,deg, is consistent with the value of $\theta_i=45\pm5\degr$ derived through geometric modeling of the narrow-line region as a bicone  \citep{das05,fischer13}.  And with an opening angle of $\theta_o=56.6^{+15.8}_{-14.3}$\,deg, the models suggest that, as observers, we are just able to peer over the edge of the BLR structure and view the innermost central engine.  This interpretation is supported by observations of X-ray variability in NGC\,4151 that seem to be caused by eclipsing material associated with the BLR traversing our line of sight \citep{puccetti07,wang10}.  

\begin{deluxetable}{lcl}
\renewcommand{\arraystretch}{1.1}
\tablecolumns{3}
\tablewidth{0pt}
\tablecaption{Direct Measurements of \mbh\ for NGC\,4151}
\tablehead{
\colhead{Method} & 
\colhead{\mbh\ ($10^{7}$ M$_{\odot}$)} & 
\colhead{Reference}
}
\startdata
RM Modeling & $1.66^{+0.48}_{-0.34}$ & This Work \\
SD Modeling & $0.25-3.0$             & \citet{roberts21} \\
%SD Modeling & $4.27^{+1.31}_{-1.31}$ & \citet{onken14}\\%
GD Modeling & $3.6^{+0.9}_{-2.6}$    & \citet{hicks08}\\
H$\beta$ RM & $4.45^{+0.79}_{-0.56}$ & \citet{bentz06b}\\
H$\beta$ RM & $2.41^{+0.18}_{-0.21}$ & \citet{derosa18}
\label{tab:mbh}
\enddata

\tablecomments{Masses from stellar dynamical (SD) and gas dynamical (GD) modeling assume a galaxy distance of 15.8 Mpc. Masses from H$\beta$ RM assume $\langle f \rangle = 4.82$.}
\end{deluxetable}

In Table~\ref{tab:mbh} we list all published direct black hole mass measurements for NGC\,4151, including the black hole mass that we present here from modeling of velocity-resolved RM data of $\log M_{\rm BH}/M_{\odot} = 7.22^{+0.11}_{-0.10}$ or $M_{\rm BH}=1.66^{+0.48}_{-0.34}\times10^7$\,\msun. \citet{roberts21} carried out a new stellar dynamical modeling analysis of the nuclear stellar kinematics presented by \citet{onken14} and determined that the models preferred a black hole mass of $M_{\rm BH}=0.25-3.0\times10^7$\,\msun, which agrees very well with the value we present here.   The mass based on gas dynamical modeling presented by \citet{hicks08} was adjusted to a galaxy distance of $D=15.8\pm0.4$\,Mpc as determined from \hst\ observations of Cepheid stars in NGC\,4151 \citep{yuan20}, and also agrees with the mass we present here as well as the mass from stellar dynamical modeling.  

Reverberation analyses of NGC\,4151 that adopt a population-average scale factor of $\langle f \rangle = 4.82\pm1.67$  \citep{batiste17b} and are listed in Table~\ref{tab:mbh} as `H$\beta$ RM', somewhat overestimate the black hole mass.  Based on the inclination angles preferred by the best-fit BLR models, this effect is to be expected.  Values of $\langle f \rangle$ range from 2.8 \citep{graham11} to 5.5 \citep{onken04} in the literature depending on the exact sample and the analysis methods employed, with most investigations settling on values of $\sim 4-5$.  Given the factor of $1/\sin \theta_i$ between the observed velocities along the line of sight and the true velocities, values of $\langle f \rangle \approx 4-5$ suggest that most Seyferts in the reverberation sample are viewed at inclinations of $25\degr-30\degr$.  An inclination angle of $\sim58\degr$, such as the models suggest for NGC\,4151, would require a smaller than average scale factor to accurately calibrate the black hole mass.  With the black hole mass constraints from the models presented here and the mean time delay and line width for H$\beta$ that were reported by \citet{bentz06b}, we can infer a specific value of $f=1.8^{+0.6}_{-0.4}$ for NGC\,4151.   \citet{williams18} explored the relationship between inclination angles and predicted scale factors for a sample of AGNs with modeling results from {\tt CARAMEL}. Their linear regression results predict an individual scale factor of $\log f = -0.44^{+1.22}_{-1.23}$ or $f=0.36^{+5.65}_{-0.34}$ for an inclination angle of $\theta_i=58.1^{+8.4}_{-9.6}$\,deg.  A similar analysis by \citet{villafana22b} with an expanded sample, including the results we present here for NGC\,4151, predicts the same individual scale factor but with somewhat smaller uncertainties, $\log f=	-0.44^{+0.73}_{-0.74}$ or $f=0.36^{+1.57}_{-0.29}$.  Both studies agree with the value that we infer here.  We note that if a scale factor of $f=1.8$ is adopted for the reverberation analysis of \citet{derosa18}, the derived mass is somewhat lower than we find here, but agrees within $2\sigma$.  It is likely that the scale factor associated with a specific AGN may change as a function of time as the detailed geometry and kinematics in the BLR change on a dynamical timescale (a few years at the location of the H$\beta$-emitting BLR in Seyferts) and respond to large-scale changes in the ionizing flux.  With 7 years passing between the 2005 observational campaign analyzed here and the 2012 program presented by \citet{derosa18}, BLR structural changes would not be unexpected.  Modeling of the 2012 data is currently in progress and may provide interesting insights into the time evolution of the structure of the BLR in NGC\,4151 (Robinson et al., in prep). 

Finally, with the successful launch of JWST in December 2021, new observations of the nuclear stellar kinematics in NGC\,4151 will soon be collected with NIRSpec as part of an Early Release Science program (ERS 1364, PI Bentz).  NIRSpec is expected to provide some crucial advantages over AO-assisted ground-based observations, such as the NIFS observations analyzed by \citet{roberts21}.  With its stable and diffraction-limited PSF and significantly lower backgrounds, stellar kinematics measured with NIRSpec may allow for tighter constraints to be placed on the black hole mass through stellar dynamical modeling, allowing further exploration of the consistency in masses derived from  independent black hole mass measurement techniques.

\section{Summary}

We have reanalyzed the 2005 monitoring data for NGC\,4151, supplemented with additional measurements from the literature, and carried out an exploration of models for the full velocity-resolved BLR response to continuum variations.  The modeling results find that the BLR is well represented by a very thick disk with an opening angle ($\theta_o\approx57\degr$) that is similar to the inclination angle ($\theta_i\approx58\degr$).  The inclination angle is consistent with the value derived from bicone modeling of the narrow line region, and the similarity of the opening angle and inclination angle suggests that our line of sight to the innermost central engine is just barely free of obstruction from the BLR and dusty torus.  This is an intriguing consideration since previous studies suggest that BLR gas has been observed to temporarily eclipse the central X-ray source along our sight line to NGC\,4151. The kinematics of the BLR gas are found to be dominated by eccentric but bound orbits, with $\sim10$\% of the orbits showing a preference for near-circular motions.  With the geometry and kinematics of the BLR constrained, the models provide an independent and direct black hole mass measurement of $\log M_{\rm BH}/M_{\odot} = 7.22^{+0.11}_{-0.10}$ or $M_{\rm BH}=1.66^{+0.48}_{-0.34}\times10^7$\,\msun, which is in excellent agreement with black hole masses determined from stellar dynamical modeling and gas dynamical modeling of NGC\,4151.

\begin{acknowledgements}
We thank the anonymous referee for helpful suggestions that improved the presentation of this work.
MCB gratefully acknowledges support from the NSF through grant AST-2009230. TT and PRW acknowledge support by the Packard Foundation through a Packard Research Fellowship to TT and from NSF through grant NSF-AST-1907208.

\end{acknowledgements}

%%% https://journals.aas.org/facility-keywords/

\facility{McGraw-Hill (Boller and Chivens CCD Spectrograph)}

\software{ULySS \citep{koleva11}, CARAMEL \citep{pancoast14a} }

%\bibliography{mbentz}{}

\begin{thebibliography}{}
\expandafter\ifx\csname natexlab\endcsname\relax\def\natexlab#1{#1}\fi
\providecommand{\url}[1]{\href{#1}{#1}}
\providecommand{\dodoi}[1]{doi:~\href{http://doi.org/#1}{\nolinkurl{#1}}}
\providecommand{\doeprint}[1]{\href{http://ascl.net/#1}{\nolinkurl{http://ascl.net/#1}}}
\providecommand{\doarXiv}[1]{\href{https://arxiv.org/abs/#1}{\nolinkurl{https://arxiv.org/abs/#1}}}

\bibitem[{{Anderson} {et~al.}(2021){Anderson}, {Baron}, \&
  {Bentz}}]{anderson21}
{Anderson}, M.~D., {Baron}, F., \& {Bentz}, M.~C. 2021, \mnras, 505, 2903,
  \dodoi{10.1093/mnras/stab1394}

\bibitem[{{Antonucci} \& {Cohen}(1983)}]{antonucci83}
{Antonucci}, R.~R.~J., \& {Cohen}, R.~D. 1983, \apj, 271, 564,
  \dodoi{10.1086/161223}

\bibitem[{{Bahcall} {et~al.}(1972){Bahcall}, {Kozlovsky}, \&
  {Salpeter}}]{bahcall72}
{Bahcall}, J.~N., {Kozlovsky}, B.-Z., \& {Salpeter}, E.~E. 1972, \apj, 171,
  467, \dodoi{10.1086/151300}

\bibitem[{{Baldwin} {et~al.}(1995){Baldwin}, {Ferland}, {Korista}, \&
  {Verner}}]{baldwin95}
{Baldwin}, J., {Ferland}, G., {Korista}, K., \& {Verner}, D. 1995, \apjl, 455,
  L119, \dodoi{10.1086/309827}

\bibitem[{{Batiste} {et~al.}(2017){Batiste}, {Bentz}, {Raimundo},
  {Vestergaard}, \& {Onken}}]{batiste17b}
{Batiste}, M., {Bentz}, M.~C., {Raimundo}, S.~I., {Vestergaard}, M., \&
  {Onken}, C.~A. 2017, \apjl, 838, L10, \dodoi{10.3847/2041-8213/aa6571}

\bibitem[{{Bentz} {et~al.}(2010){Bentz}, {Horne}, {Barth}, {et~al.}}]{bentz10b}
{Bentz}, M.~C., {Horne}, K., {Barth}, A.~J., {et~al.} 2010, \apjl, 720, L46,
  \dodoi{10.1088/2041-8205/720/1/L46}

\bibitem[{{Bentz} {et~al.}(2009){Bentz}, {Walsh}, {Barth}, {et~al.}}]{bentz09c}
{Bentz}, M.~C., {Walsh}, J.~L., {Barth}, A.~J., {et~al.} 2009, \apj, 705, 199,
  \dodoi{10.1088/0004-637X/705/1/199}

\bibitem[{{Bentz} {et~al.}(2021){Bentz}, {Williams}, {Street}, {Onken},
  {Valluri}, \& {Treu}}]{bentz21c}
{Bentz}, M.~C., {Williams}, P.~R., {Street}, R., {et~al.} 2021, \apj, 920, 112,
  \dodoi{10.3847/1538-4357/ac19af}

\bibitem[{{Bentz} {et~al.}(2006){Bentz}, {Denney}, {Cackett}, {Dietrich},
  {Fogel}, {Ghosh}, {Horne}, {Kuehn}, {Minezaki}, {Onken}, {Peterson}, {Pogge},
  {Pronik}, {Richstone}, {Sergeev}, {Vestergaard}, {Walker}, \&
  {Yoshii}}]{bentz06b}
{Bentz}, M.~C., {Denney}, K.~D., {Cackett}, E.~M., {et~al.} 2006, \apj, 651,
  775, \dodoi{10.1086/507417}

\bibitem[{{Bentz} {et~al.}(2008){Bentz}, {Walsh}, {Barth}, {Baliber},
  {Bennert}, {Canalizo}, {Filippenko}, {Ganeshalingam}, {Gates}, {Greene},
  {Hidas}, {Hiner}, {Lee}, {Li}, {Malkan}, {Minezaki}, {Serduke}, {Shiode},
  {Silverman}, {Steele}, {Stern}, {Street}, {Thornton}, {Treu}, {Wang}, {Woo},
  \& {Yoshii}}]{bentz08}
{Bentz}, M.~C., {Walsh}, J.~L., {Barth}, A.~J., {et~al.} 2008, \apjl, 689, L21,
  \dodoi{10.1086/595719}

\bibitem[{{Blandford} \& {McKee}(1982)}]{blandford82}
{Blandford}, R.~D., \& {McKee}, C.~F. 1982, \apj, 255, 419,
  \dodoi{10.1086/159843}

\bibitem[{{Bochkarev}(1984)}]{bochkarev84}
{Bochkarev}, N.~G. 1984, Soviet Astronomy Letters, 10, 239

\bibitem[{Brewer \& Foreman-Mackey(2018)}]{brewer18}
Brewer, B.~J., \& Foreman-Mackey, D. 2018, Journal of Statistical Software,
  Articles, 86, 1, \dodoi{10.18637/jss.v086.i07}

\bibitem[{{Brewer} {et~al.}(2011){Brewer}, {Treu}, {Pancoast}, {Barth},
  {Bennert}, {Bentz}, {Filippenko}, {Greene}, {Malkan}, \& {Woo}}]{brewer11}
{Brewer}, B.~J., {Treu}, T., {Pancoast}, A., {et~al.} 2011, \apjl, 733, L33,
  \dodoi{10.1088/2041-8205/733/2/L33}

\bibitem[{Cackett {et~al.}(2021)Cackett, Bentz, \& Kara}]{cackett21}
Cackett, E.~M., Bentz, M.~C., \& Kara, E. 2021, iScience, 24, 102557,
  \dodoi{https://doi.org/10.1016/j.isci.2021.102557}

\bibitem[{{Campbell} \& {Moore}(1918)}]{campbell18}
{Campbell}, W.~W., \& {Moore}, J.~H. 1918, Publications of Lick Observatory,
  13, 75

\bibitem[{{Cherepashchuk} \& {Lyutyi}(1973)}]{cherepashchuk73}
{Cherepashchuk}, A.~M., \& {Lyutyi}, V.~M. 1973, \aplett, 13, 165

\bibitem[{{Clavel} {et~al.}(1990){Clavel}, {Boksenberg}, {Bromage}, {Elvius},
  {Penston}, {Perola}, {Santos-Leo}, {Snijders}, \& {Ulrich}}]{clavel90}
{Clavel}, J., {Boksenberg}, A., {Bromage}, G.~E., {et~al.} 1990, \mnras, 246,
  668

\bibitem[{{Das} {et~al.}(2005){Das}, {Crenshaw}, {Hutchings}, {Deo}, {Kraemer},
  {Gull}, {Kaiser}, {Nelson}, \& {Weistrop}}]{das05}
{Das}, V., {Crenshaw}, D.~M., {Hutchings}, J.~B., {et~al.} 2005, \aj, 130, 945,
  \dodoi{10.1086/432255}

\bibitem[{{De Rosa} {et~al.}(2018){De Rosa}, {Fausnaugh}, {Grier}, {Peterson},
  {Denney}, {Horne}, {Bentz}, {Ciroi}, {Dalla Bont{\`a}}, {Joner}, {Kaspi},
  {Kochanek}, {Pogge}, {Sergeev}, {Vestergaard}, {Adams}, {Antognini}, {Araya
  Salvo}, {Armstrong}, {Bae}, {Barth}, {Beatty}, {Bhattacharjee}, {Borman},
  {Boroson}, {Bottorff}, {Brown}, {Brown}, {Brotherton}, {Coker}, {Clanton},
  {Cracco}, {Crawford}, {Croxall}, {Eftekharzadeh}, {Eracleous}, {Fiorenza},
  {Frassati}, {Hawkins}, {Henderson}, {Holoien}, {Hutchison}, {Kellar},
  {Kilerci-Eser}, {Kim}, {King}, {La Mura}, {Laney}, {Li}, {Lochhaas}, {Ma},
  {MacInnis}, {Manne-Nicholas}, {Mason}, {McGraw}, {Mogren}, {Montouri},
  {Moody}, {Mosquera}, {Mudd}, {Musso}, {Nazarov}, {Nguyen}, {Ochner},
  {Okhmat}, {Onken}, {Ou-Yang}, {Pancoast}, {Pei}, {Penny}, {Poleski},
  {Portaluri}, {Prieto}, {Price-Whelan}, {Pulatova}, {Rafter}, {Roettenbacher},
  {Romero-Colmenero}, {Runnoe}, {Schimoia}, {Shappee}, {Sherf}, {Simonian},
  {Siviero}, {Skowron}, {Skowron}, {Somers}, {Spencer}, {Starkey}, {Stevens},
  {Stoll}, {Tamajo}, {Tayar}, {van Saders}, {Valenti}, {Villanueva},
  {Villforth}, {Weiss}, {Winkler}, {Zastrow}, {Zhu}, \& {Zu}}]{derosa18}
{De Rosa}, G., {Fausnaugh}, M.~M., {Grier}, C.~J., {et~al.} 2018, \apj, 866,
  133, \dodoi{10.3847/1538-4357/aadd11}

\bibitem[{{Denney} {et~al.}(2006){Denney}, {Bentz}, {Peterson}, {Pogge},
  {Cackett}, {Dietrich}, {Fogel}, {Ghosh}, {Horne}, {Kuehn}, {Minezaki},
  {Onken}, {Pronik}, {Richstone}, {Sergeev}, {Vestergaard}, {Walker}, \&
  {Yoshii}}]{denney06}
{Denney}, K.~D., {Bentz}, M.~C., {Peterson}, B.~M., {et~al.} 2006, \apj, 653,
  152, \dodoi{10.1086/508533}

\bibitem[{{Denney} {et~al.}(2009){Denney}, {Peterson}, {Pogge}, {Adair},
  {Atlee}, {Au-Yong}, {Bentz}, {Bird}, {Brokofsky}, {Chisholm}, {Comins},
  {Dietrich}, {Doroshenko}, {Eastman}, {Efimov}, {Ewald}, {Ferbey}, {Gaskell},
  {Hedrick}, {Jackson}, {Klimanov}, {Klimek}, {Kruse}, {Lad{\'e}route}, {Lamb},
  {Leighly}, {Minezaki}, {Nazarov}, {Onken}, {Petersen}, {Peterson},
  {Poindexter}, {Sakata}, {Schlesinger}, {Sergeev}, {Skolski}, {Stieglitz},
  {Tobin}, {Unterborn}, {Vestergaard}, {Watkins}, {Watson}, \&
  {Yoshii}}]{denney09c}
{Denney}, K.~D., {Peterson}, B.~M., {Pogge}, R.~W., {et~al.} 2009, \apjl, 704,
  L80, \dodoi{10.1088/0004-637X/704/2/L80}

\bibitem[{{Event Horizon Telescope Collaboration} {et~al.}(2019){Event Horizon
  Telescope Collaboration}, {Akiyama}, {Alberdi}, {et~al.}}]{eht19}
{Event Horizon Telescope Collaboration}, {Akiyama}, K., {Alberdi}, A., {et~al.}
  2019, \apjl, 875, L6, \dodoi{10.3847/2041-8213/ab1141}

\bibitem[{{Fath}(1909)}]{fath09}
{Fath}, E.~A. 1909, Lick Observatory Bull., 149, 71,
  \dodoi{10.5479/ADS/bib/1909LicOB.5.71F}

\bibitem[{{Fischer} {et~al.}(2013){Fischer}, {Crenshaw}, {Kraemer}, \&
  {Schmitt}}]{fischer13}
{Fischer}, T.~C., {Crenshaw}, D.~M., {Kraemer}, S.~B., \& {Schmitt}, H.~R.
  2013, \apjs, 209, 1, \dodoi{10.1088/0067-0049/209/1/1}

\bibitem[{{Genzel} {et~al.}(2000){Genzel}, {Pichon}, {Eckart}, {Gerhard}, \&
  {Ott}}]{genzel00}
{Genzel}, R., {Pichon}, C., {Eckart}, A., {Gerhard}, O.~E., \& {Ott}, T. 2000,
  \mnras, 317, 348, \dodoi{10.1046/j.1365-8711.2000.03582.x}

\bibitem[{{Ghez} {et~al.}(2000){Ghez}, {Morris}, {Becklin}, {Tanner}, \&
  {Kremenek}}]{ghez00}
{Ghez}, A.~M., {Morris}, M., {Becklin}, E.~E., {Tanner}, A., \& {Kremenek}, T.
  2000, \nat, 407, 349, \dodoi{10.1038/407349a}

\bibitem[{{Ghez} {et~al.}(2008){Ghez}, {Salim}, {Weinberg}, {Lu}, {Do}, {Dunn},
  {Matthews}, {Morris}, {Yelda}, {Becklin}, {Kremenek}, {Milosavljevic}, \&
  {Naiman}}]{ghez08}
{Ghez}, A.~M., {Salim}, S., {Weinberg}, N.~N., {et~al.} 2008, \apj, 689, 1044,
  \dodoi{10.1086/592738}

\bibitem[{{Graham} {et~al.}(2011){Graham}, {Onken}, {Athanassoula}, \&
  {Combes}}]{graham11}
{Graham}, A.~W., {Onken}, C.~A., {Athanassoula}, E., \& {Combes}, F. 2011,
  \mnras, 412, 2211, \dodoi{10.1111/j.1365-2966.2010.18045.x}

\bibitem[{{Gravity Collaboration} {et~al.}(2018){Gravity Collaboration},
  {Sturm}, {Dexter}, {Pfuhl}, {Stock}, {Davies}, {Lutz}, {Cl{\'e}net},
  {Eckart}, {Eisenhauer}, {Genzel}, {Gratadour}, {H{\"o}nig}, {Kishimoto},
  {Lacour}, {Millour}, {Netzer}, {Perrin}, {Peterson}, {Petrucci}, {Rouan},
  {Waisberg}, {Woillez}, {Amorim}, {Brandner}, {F{\"o}rster Schreiber},
  {Garcia}, {Gillessen}, {Ott}, {Paumard}, {Perraut}, {Scheithauer},
  {Straubmeier}, {Tacconi}, \& {Widmann}}]{gravity18}
{Gravity Collaboration}, {Sturm}, E., {Dexter}, J., {et~al.} 2018, \nat, 563,
  657, \dodoi{10.1038/s41586-018-0731-9}

\bibitem[{{Gravity Collaboration} {et~al.}(2020){Gravity Collaboration},
  {Amorim}, {Baub{\"o}ck}, {Brandner}, {Cl{\'e}net}, {Davies}, {de Zeeuw},
  {Dexter}, {Eckart}, {Eisenhauer}, {F{\"o}rster Schreiber}, {Gao}, {Garcia},
  {Genzel}, {Gillessen}, {Gratadour}, {H{\"o}nig}, {Kishimoto}, {Lacour},
  {Lutz}, {Millour}, {Netzer}, {Ott}, {Paumard}, {Perraut}, {Perrin},
  {Peterson}, {Petrucci}, {Pfuhl}, {Prieto}, {Rouan}, {Shangguan}, {Shimizu},
  {Schartmann}, {Stadler}, {Sternberg}, {Straub}, {Straubmeier}, {Sturm},
  {Tacconi}, {Tristram}, {Vermot}, {von Fellenberg}, {Waisberg}, {Widmann}, \&
  {Woillez}}]{gravity20}
{Gravity Collaboration}, {Amorim}, A., {Baub{\"o}ck}, M., {et~al.} 2020, \aap,
  643, A154, \dodoi{10.1051/0004-6361/202039067}

\bibitem[{{Grier} {et~al.}(2017){Grier}, {Pancoast}, {Barth}, {Fausnaugh},
  {Brewer}, {Treu}, \& {Peterson}}]{grier17}
{Grier}, C.~J., {Pancoast}, A., {Barth}, A.~J., {et~al.} 2017, \apj, 849, 146,
  \dodoi{10.3847/1538-4357/aa901b}

\bibitem[{{Grier} {et~al.}(2012){Grier}, {Peterson}, {Pogge},
  {et~al.}}]{grier12b}
{Grier}, C.~J., {Peterson}, B.~M., {Pogge}, R.~W., {et~al.} 2012, \apj, 755,
  60, \dodoi{10.1088/0004-637X/755/1/60}

\bibitem[{{Grier} {et~al.}(2019){Grier}, {Shen}, {Horne}, {Brandt}, {Trump},
  {Hall}, {Kinemuchi}, {Starkey}, {Schneider}, {Ho}, {Homayouni}, {I-Hsiu Li},
  {McGreer}, {Peterson}, {Bizyaev}, {Chen}, {Dawson}, {Eftekharzadeh}, {Guo},
  {Jia}, {Jiang}, {Kneib}, {Li}, {Li}, {Nie}, {Oravetz}, {Oravetz}, {Pan},
  {Petitjean}, {Ponder}, {Rogerson}, {Vivek}, {Zhang}, \& {Zou}}]{grier19}
{Grier}, C.~J., {Shen}, Y., {Horne}, K., {et~al.} 2019, \apj, 887, 38,
  \dodoi{10.3847/1538-4357/ab4ea5}

\bibitem[{{Hicks} \& {Malkan}(2008)}]{hicks08}
{Hicks}, E. K.~S., \& {Malkan}, M.~A. 2008, \apjs, 174, 31,
  \dodoi{10.1086/521650}

\bibitem[{{Hoormann} {et~al.}(2019){Hoormann}, {Martini}, {Davis}, {King},
  {Lidman}, {Mudd}, {Sharp}, {Sommer}, {Tucker}, {Yu}, {Allam}, {Asorey},
  {Avila}, {Banerji}, {Brooks}, {Buckley-Geer}, {Burke}, {Calcino}, {Carnero
  Rosell}, {Carollo}, {Carrasco Kind}, {Carretero}, {Castander}, {Childress},
  {De Vicente}, {Desai}, {Diehl}, {Doel}, {Flaugher}, {Fosalba}, {Frieman},
  {Garc{\'\i}a-Bellido}, {Gerdes}, {Gruen}, {Gutierrez}, {Hartley}, {Hinton},
  {Hollowood}, {Honscheid}, {Hoyle}, {James}, {Krause}, {Kuehn}, {Kuropatkin},
  {Lewis}, {Lima}, {Macaulay}, {Maia}, {Menanteau}, {Miller}, {Miquel},
  {M{\"o}ller}, {Plazas}, {Romer}, {Roodman}, {Sanchez}, {Scarpine},
  {Schubnell}, {Serrano}, {Sevilla-Noarbe}, {Smith}, {Smith}, {Soares-Santos},
  {Sobreira}, {Suchyta}, {Swann}, {Swanson}, {Tarle}, {Uddin}, \& {DES
  Collaboration}}]{hoormann19}
{Hoormann}, J.~K., {Martini}, P., {Davis}, T.~M., {et~al.} 2019, \mnras, 487,
  3650, \dodoi{10.1093/mnras/stz1539}

\bibitem[{{Horne}(1994)}]{horne94}
{Horne}, K. 1994, in Astronomical Society of the Pacific Conference Series,
  Vol.~69, Reverberation Mapping of the Broad-Line Region in Active Galactic
  Nuclei, ed. P.~M. {Gondhalekar}, K.~{Horne}, \& B.~M. {Peterson}, 23

\bibitem[{{Horne} {et~al.}(2021){Horne}, {De Rosa}, {Peterson}, {Barth}, {Ely},
  {Fausnaugh}, {Kriss}, {Pei}, {Bentz}, {Cackett}, {Edelson}, {Eracleous},
  {Goad}, {Grier}, {Kaastra}, {Kochanek}, {Krongold}, {Mathur}, {Netzer},
  {Proga}, {Tejos}, {Vestergaard}, {Villforth}, {Adams}, {Anderson},
  {Ar{\'e}valo}, {Beatty}, {Bennert}, {Bigley}, {Bisogni}, {Borman}, {Boroson},
  {Bottorff}, {Brandt}, {Breeveld}, {Brotherton}, {Brown}, {Brown}, {Canalizo},
  {Carini}, {Clubb}, {Comerford}, {Corsini}, {Crenshaw}, {Croft}, {Croxall},
  {Dalla Bont{\`a}}, {Deason}, {Dehghanian}, {De Lorenzo-C{\'a}ceres},
  {Denney}, {Dietrich}, {Done}, {Efimova}, {Evans}, {Ferland}, {Filippenko},
  {Flatland}, {Fox}, {Gardner}, {Gates}, {Gehrels}, {Geier}, {Gelbord},
  {Gonzalez}, {Gorjian}, {Greene}, {Grupe}, {Gupta}, {Hall}, {Henderson},
  {Hicks}, {Holmbeck}, {Holoien}, {Hutchison}, {Im}, {Jensen}, {Johnson},
  {Joner}, {Jones}, {Kaspi}, {Kelly}, {Kennea}, {Kim}, {Kim}, {Kim}, {King},
  {Klimanov}, {Korista}, {Lau}, {Lee}, {Leonard}, {Li}, {Lira}, {Lochhaas},
  {Ma}, {MacInnis}, {Malkan}, {Manne-Nicholas}, {Mauerhan}, {McGurk},
  {McHardy}, {Montuori}, {Morelli}, {Mosquera}, {Mudd},
  {M{\"u}ller-S{\'a}nchez}, {Nazarov}, {Norris}, {Nousek}, {Nguyen}, {Ochner},
  {Okhmat}, {Pancoast}, {Papadakis}, {Parks}, {Penny}, {Pizzella}, {Pogge},
  {Poleski}, {Pott}, {Rafter}, {Rix}, {Runnoe}, {Saylor}, {Schimoia},
  {Schn{\"u}lle}, {Scott}, {Sergeev}, {Shappee}, {Shivvers}, {Siegel},
  {Simonian}, {Siviero}, {Skielboe}, {Somers}, {Spencer}, {Starkey}, {Stevens},
  {Sung}, {Tayar}, {Treu}, {Turner}, {Uttley}, {Van Saders}, {Vican},
  {Villanueva}, {Weiss}, {Woo}, {Yan}, {Young}, {Yuk}, {Zheng}, {Zhu}, \&
  {Zu}}]{horne21}
{Horne}, K., {De Rosa}, G., {Peterson}, B.~M., {et~al.} 2021, \apj, 907, 76,
  \dodoi{10.3847/1538-4357/abce60}

\bibitem[{{Kaspi} {et~al.}(2007){Kaspi}, {Brandt}, {Maoz}, {Netzer},
  {Schneider}, \& {Shemmer}}]{kaspi07}
{Kaspi}, S., {Brandt}, W.~N., {Maoz}, D., {et~al.} 2007, \apj, 659, 997,
  \dodoi{10.1086/512094}

\bibitem[{{Koleva} {et~al.}(2009){Koleva}, {Prugniel}, {Bouchard}, \&
  {Wu}}]{koleva09}
{Koleva}, M., {Prugniel}, P., {Bouchard}, A., \& {Wu}, Y. 2009, \aap, 501,
  1269, \dodoi{10.1051/0004-6361/200811467}

\bibitem[{{Koleva} {et~al.}(2011){Koleva}, {Prugniel}, {Bouchard}, \&
  {Wu}}]{koleva11}
---. 2011, {ULySS: A Full Spectrum Fitting Package}.
\newblock \doeprint{1104.007}

\bibitem[{{Kormendy} \& {Ho}(2013)}]{kormendy13}
{Kormendy}, J., \& {Ho}, L.~C. 2013, \araa, 51, 511,
  \dodoi{10.1146/annurev-astro-082708-101811}

\bibitem[{{Koshida} {et~al.}(2014){Koshida}, {Minezaki}, {Yoshii}, {Kobayashi},
  {Sakata}, {Sugawara}, {Enya}, {Suganuma}, {Tomita}, {Aoki}, \&
  {Peterson}}]{koshida14}
{Koshida}, S., {Minezaki}, T., {Yoshii}, Y., {et~al.} 2014, \apj, 788, 159,
  \dodoi{10.1088/0004-637X/788/2/159}

\bibitem[{{Mangham} {et~al.}(2019){Mangham}, {Knigge}, {Williams}, {Horne},
  {Pancoast}, {Matthews}, {Long}, {Sim}, \& {Higginbottom}}]{mangham19}
{Mangham}, S.~W., {Knigge}, C., {Williams}, P., {et~al.} 2019, \mnras, 488,
  2780, \dodoi{10.1093/mnras/stz1713}

\bibitem[{{Maoz} {et~al.}(1991){Maoz}, {Netzer}, {Mazeh}, {Beck}, {Almoznino},
  {Leibowitz}, {Brosch}, {Mendelson}, \& {Laor}}]{maoz91}
{Maoz}, D., {Netzer}, H., {Mazeh}, T., {et~al.} 1991, \apj, 367, 493,
  \dodoi{10.1086/169646}

\bibitem[{{Merritt}(2022)}]{merritt22}
{Merritt}, R. 2022, PhD thesis, Georgia State University.
\newblock \url{https://scholarworks.gsu.edu/phy_astr_diss/144}

\bibitem[{{Miyoshi} {et~al.}(1995){Miyoshi}, {Moran}, {Herrnstein},
  {Greenhill}, {Nakai}, {Diamond}, \& {Inoue}}]{miyoshi95}
{Miyoshi}, M., {Moran}, J., {Herrnstein}, J., {et~al.} 1995, \nat, 373, 127,
  \dodoi{10.1038/373127a0}

\bibitem[{{Oknyanskij} {et~al.}(2016){Oknyanskij}, {Metlova}, {Huseynov},
  {Guo}, \& {Lyuty}}]{oknyanskij16}
{Oknyanskij}, V.~L., {Metlova}, N.~V., {Huseynov}, N.~A., {Guo}, D.-F., \&
  {Lyuty}, V.~M. 2016, Odessa Astronomical Publications, 29, 95,
  \dodoi{10.18524/1810-4215.2016.29.85058}

\bibitem[{{Onken} {et~al.}(2004){Onken}, {Ferrarese}, {Merritt}, {Peterson},
  {Pogge}, {Vestergaard}, \& {Wandel}}]{onken04}
{Onken}, C.~A., {Ferrarese}, L., {Merritt}, D., {et~al.} 2004, \apj, 615, 645,
  \dodoi{10.1086/424655}

\bibitem[{{Onken} {et~al.}(2014){Onken}, {Valluri}, {Brown},
  {et~al.}}]{onken14}
{Onken}, C.~A., {Valluri}, M., {Brown}, J.~S., {et~al.} 2014, \apj, 791, 37,
  \dodoi{10.1088/0004-637X/791/1/37}

\bibitem[{{Pancoast} {et~al.}(2011){Pancoast}, {Brewer}, \&
  {Treu}}]{pancoast11}
{Pancoast}, A., {Brewer}, B.~J., \& {Treu}, T. 2011, \apj, 730, 139,
  \dodoi{10.1088/0004-637X/730/2/139}

\bibitem[{{Pancoast} {et~al.}(2014{\natexlab{a}}){Pancoast}, {Brewer}, \&
  {Treu}}]{pancoast14a}
---. 2014{\natexlab{a}}, \mnras, 445, 3055, \dodoi{10.1093/mnras/stu1809}

\bibitem[{{Pancoast} {et~al.}(2014{\natexlab{b}}){Pancoast}, {Brewer}, {Treu},
  {Park}, {Barth}, {Bentz}, \& {Woo}}]{pancoast14b}
{Pancoast}, A., {Brewer}, B.~J., {Treu}, T., {et~al.} 2014{\natexlab{b}},
  \mnras, 445, 3073, \dodoi{10.1093/mnras/stu1419}

\bibitem[{{Pancoast} {et~al.}(2018){Pancoast}, {Barth}, {Horne}, {Treu},
  {Brewer}, {Bennert}, {Canalizo}, {Gates}, {Li}, {Malkan}, {Sand}, {Schmidt},
  {Valenti}, {Woo}, {Clubb}, {Cooper}, {Crawford}, {H{\"o}nig}, {Joner},
  {Kandrashoff}, {Lazarova}, {Nierenberg}, {Romero-Colmenero}, {Son},
  {Tollerud}, {Walsh}, \& {Winkler}}]{pancoast18}
{Pancoast}, A., {Barth}, A.~J., {Horne}, K., {et~al.} 2018, \apj, 856, 108,
  \dodoi{10.3847/1538-4357/aab3c6}

\bibitem[{{Peterson}(2010)}]{peterson10}
{Peterson}, B.~M. 2010, in IAU Symposium, Vol. 267, IAU Symposium, ed. B.~M.
  {Peterson}, R.~S. {Somerville}, \& T.~{Storchi-Bergmann}, 151--160,
  \dodoi{10.1017/S1743921310006095}

\bibitem[{{Peterson} \& {Cota}(1988)}]{peterson88}
{Peterson}, B.~M., \& {Cota}, S.~A. 1988, \apj, 330, 111,
  \dodoi{10.1086/166459}

\bibitem[{{Peterson} {et~al.}(1985){Peterson}, {Meyers}, {Capriotti}, {Foltz},
  {Wilkes}, \& {Miller}}]{peterson85}
{Peterson}, B.~M., {Meyers}, K.~A., {Capriotti}, E.~R., {et~al.} 1985, \apj,
  292, 164, \dodoi{10.1086/163142}

\bibitem[{{Peterson} \& {Wandel}(1999)}]{peterson99}
{Peterson}, B.~M., \& {Wandel}, A. 1999, \apjl, 521, L95,
  \dodoi{10.1086/312190}

\bibitem[{{Peterson} \& {Wandel}(2000)}]{peterson00a}
---. 2000, \apjl, 540, L13, \dodoi{10.1086/312862}

\bibitem[{{Peterson} {et~al.}(1991){Peterson}, {Balonek}, {Barker}, {Bechtold},
  {Bertram}, {Bochkarev}, {Bolte}, {Bond}, {Boroson}, {Carini}, {Carone},
  {Christensen}, {Clements}, {Cochran}, {Cohen}, {Crampton}, {Dietrich},
  {Elvis}, {Ferguson}, {Filippenko}, {Fricke}, {Gaskell}, {Halpern}, {Huchra},
  {Hutchings}, {Kollatschny}, {Koratkar}, {Korista}, {Krolik}, {Lame}, {Laor},
  {Leacock}, {MacAlpine}, {Malkan}, {Maoz}, {Miller}, {Morris}, {Netzer},
  {Oliveira}, {Penfold}, {Penston}, {Perez}, {Pogge}, {Richmond}, {Romanishin},
  {Rosenblatt}, {Saddlemyer}, {Sadun}, {Sawyer}, {Shields}, {Shapovalova},
  {Smith}, {Smith}, {Smith}, {Sun}, {Thiele}, {Turner}, {Veilleux}, {Wagner},
  {Weymann}, {Wilkes}, {Wills}, {Wills}, \& {Younger}}]{peterson91}
{Peterson}, B.~M., {Balonek}, T.~J., {Barker}, E.~S., {et~al.} 1991, \apj, 368,
  119, \dodoi{10.1086/169675}

\bibitem[{{Puccetti} {et~al.}(2007){Puccetti}, {Fiore}, {Risaliti}, {Capalbi},
  {Elvis}, \& {Nicastro}}]{puccetti07}
{Puccetti}, S., {Fiore}, F., {Risaliti}, G., {et~al.} 2007, \mnras, 377, 607,
  \dodoi{10.1111/j.1365-2966.2007.11634.x}

\bibitem[{{Roberts} {et~al.}(2021){Roberts}, {Bentz}, {Vasiliev}, {Valluri}, \&
  {Onken}}]{roberts21}
{Roberts}, C.~A., {Bentz}, M.~C., {Vasiliev}, E., {Valluri}, M., \& {Onken},
  C.~A. 2021, \apj, 916, 25, \dodoi{10.3847/1538-4357/ac05b6}

\bibitem[{{Roberts} \& {Rumstay}(2012)}]{roberts12}
{Roberts}, C.~A., \& {Rumstay}, K.~R. 2012, Journal of the Southeastern
  Association for Research in Astronomy, 6, 47

\bibitem[{{Seyfert}(1943)}]{seyfert43}
{Seyfert}, C.~K. 1943, ApJ, 97, 28, \dodoi{10.1086/144488}

\bibitem[{{Shen} {et~al.}(2011){Shen}, {Richards}, {Strauss}, {Hall},
  {Schneider}, {Snedden}, {Bizyaev}, {Brewington}, {Malanushenko},
  {Malanushenko}, {Oravetz}, {Pan}, \& {Simmons}}]{shen11}
{Shen}, Y., {Richards}, G.~T., {Strauss}, M.~A., {et~al.} 2011, \apjs, 194, 45,
  \dodoi{10.1088/0067-0049/194/2/45}

\bibitem[{{Skielboe} {et~al.}(2015){Skielboe}, {Pancoast}, {Treu}, {Park},
  {Barth}, \& {Bentz}}]{skielboe15}
{Skielboe}, A., {Pancoast}, A., {Treu}, T., {et~al.} 2015, \mnras, 454, 144,
  \dodoi{10.1093/mnras/stv1917}

\bibitem[{{Ulrich} {et~al.}(1991){Ulrich}, {Boksenberg}, {Bromage}, {Clavel},
  {Elvius}, {Penston}, {Perola}, \& {Snijders}}]{ulrich91}
{Ulrich}, M.~H., {Boksenberg}, A., {Bromage}, G.~E., {et~al.} 1991, \apj, 382,
  483, \dodoi{10.1086/170735}

\bibitem[{{van Groningen} \& {Wanders}(1992)}]{vangroningen92}
{van Groningen}, E., \& {Wanders}, I. 1992, \pasp, 104, 700,
  \dodoi{10.1086/133039}

\bibitem[{{Vazdekis} {et~al.}(2010){Vazdekis}, {S{\'a}nchez-Bl{\'a}zquez},
  {Falc{\'o}n-Barroso}, {Cenarro}, {Beasley}, {Cardiel}, {Gorgas}, \&
  {Peletier}}]{vazdekis10}
{Vazdekis}, A., {S{\'a}nchez-Bl{\'a}zquez}, P., {Falc{\'o}n-Barroso}, J.,
  {et~al.} 2010, \mnras, 404, 1639, \dodoi{10.1111/j.1365-2966.2010.16407.x}

\bibitem[{{Villafa{\~n}a} {et~al.}(2022{\natexlab{a}}){Villafa{\~n}a},
  {Williams}, {Treu}, {Brewer}, {Barth}, {U}, {Bennert}, {Alexander Vogler},
  {Guo}, {Bentz}, {Canalizo}, {Filippenko}, {Gates}, {Hamann}, {Joner},
  {Malkan}, {Woo}, {Abolfathi}, {Abramson}, {Armen}, {Bae}, {Bohn}, {Boizelle},
  {Bostroem}, {Brandel}, {Brink}, {Channa}, {Cooper}, {Cosens}, {Donohue},
  {Fillingham}, {Gonz{\'a}lez-Buitrago}, {Halevi}, {Halle}, {Hood}, {Horne},
  {Chuck Horst}, {de Kouchkovsky}, {Kuhn}, {Kumar}, {Leonard}, {Loveland},
  {Manzano-King}, {McHardy}, {Michel}, {Olaes}, {Park}, {Park}, {Pei}, {Ross},
  {Runco}, {S{\'a}nchez}, {Scott}, {Sexton}, {Shin}, {Shivvers}, {Spencer},
  {Stahl}, {Stegman}, {Stomberg}, {Valenti}, {Walsh}, {Yuk}, \&
  {Zheng}}]{villafana22a}
{Villafa{\~n}a}, L., {Williams}, P.~R., {Treu}, T., {et~al.}
  2022{\natexlab{a}}, \apj, 930, 52, \dodoi{10.3847/1538-4357/ac6171}

\bibitem[{{Villafa{\~n}a} {et~al.}(2022{\natexlab{b}}){Villafa{\~n}a},
  {Williams}, {Treu}, {Brewer}, {Barth}, {U}, {Bennert}, {Vogler}, {Guo},
  {Bentz}, {Canalizo}, {Filippenko}, {Gates}, {Hamann}, {Joner}, {Malkan},
  {Woo}, {Abolfathi}, {Abramson}, {Armen}, {Bae}, {Bohn}, {Boizelle},
  {Bostroem}, {Brandel}, {Brink}, {Channa}, {Cooper}, {Cosens}, {Donohue},
  {Fillingham}, {Gonzalez-Buitrago}, {Halevi}, {Halle}, {Hood}, {Horne},
  {Horst}, {de Kouchkovsky}, {Kuhn}, {Kumar}, {Leonard}, {Loveland},
  {Manzano-King}, {McHardy}, {Michel}, {Olaes}, {Park}, {Park}, {Pei}, {Ross},
  {Runco}, {Samuel}, {Sanchez}, {Scott}, {Sexton}, {Shin}, {Shivvers},
  {Spencer}, {Stahl}, {Stegman}, {Stomberg}, {Valenti}, {Walsh}, {Yuk}, \&
  {Zheng}}]{villafana22b}
---. 2022{\natexlab{b}}, \apj, submitted

\bibitem[{{Wang} {et~al.}(2010){Wang}, {Risaliti}, {Fabbiano}, {Elvis},
  {Zezas}, \& {Karovska}}]{wang10}
{Wang}, J., {Risaliti}, G., {Fabbiano}, G., {et~al.} 2010, \apj, 714, 1497,
  \dodoi{10.1088/0004-637X/714/2/1497}

\bibitem[{{Williams} {et~al.}(2018){Williams}, {Pancoast}, {Treu},
  {et~al.}}]{williams18}
{Williams}, P.~R., {Pancoast}, A., {Treu}, T., {et~al.} 2018, \apj, 866, 75,
  \dodoi{10.3847/1538-4357/aae086}

\bibitem[{{Williams} {et~al.}(2020){Williams}, {Pancoast}, {Treu},
  {et~al.}}]{williams20}
---. 2020, \apj, 902, 74, \dodoi{10.3847/1538-4357/abbad7}

\bibitem[{{Williams} {et~al.}(2021{\natexlab{a}}){Williams}, {Treu}, {Dahle},
  {Valenti}, {Abramson}, {Barth}, {Dyrland}, {Gladders}, {Horne}, \&
  {Sharon}}]{williams21a}
{Williams}, P.~R., {Treu}, T., {Dahle}, H., {et~al.} 2021{\natexlab{a}}, \apj,
  911, 64, \dodoi{10.3847/1538-4357/abe943}

\bibitem[{{Williams} {et~al.}(2021{\natexlab{b}}){Williams}, {Treu}, {Dahle},
  {Valenti}, {Abramson}, {Barth}, {Brewer}, {Dyrland}, {Gladders}, {Horne}, \&
  {Sharon}}]{williams21b}
---. 2021{\natexlab{b}}, \apjl, 915, L9, \dodoi{10.3847/2041-8213/ac081b}

\bibitem[{{Yuan} {et~al.}(2020)}]{yuan20}
{Yuan}, W., {et~al.} 2020, ApJ, 902, 26, \dodoi{10.3847/1538-4357/abb377}

\end{thebibliography}
\bibliographystyle{aasjournal}

\end{document}